\magnification=1100

\font\bigbf=cmbx9 scaled 1500
\hfuzz 25pt
\hsize=6.1truein
\hoffset=0.2truein

\def\hi{\noindent \hangindent=2.5em}
\def\go{\mathrel{\raise.3ex\hbox{$>$}\mkern-14mu\lower0.6ex\hbox{$\sim$}}}
\def\lo{\mathrel{\raise.3ex\hbox{$<$}\mkern-14mu\lower0.6ex\hbox{$\sim$}}}
\def\eps {{\varepsilon}}

\def\rH {{\rm H}}

\centerline{\bigbf HYDROGEN PHASES ON THE SURFACE OF}
\medskip
\centerline{\bigbf A STRONGLY MAGNETIZED NEUTRON STAR}

\bigskip
\centerline{Dong Lai}
\smallskip
\centerline{\it Theoretical Astrophysics, 130-33, California Institute 
of Technology}
\centerline{\it Pasadena, CA 91125}
\centerline{E-mail: dong@tapir.caltech.edu}
\medskip
\centerline{Edwin E. Salpeter}
\smallskip
\centerline{\it Center for Radiophysics and Space Research, 
Cornell University}
\centerline{\it Ithaca, NY 14853}

\bigskip
\bigskip
\centerline{(March 1997)}
\bigskip
\bigskip
\bigskip
\centerline{\bf ABSTRACT}
\bigskip

The outermost layers of some neutron stars are likely to be dominated by
hydrogen, as a result of fast gravitational settling of heavier elements. 
These layers directly mediate thermal radiation from the stars, 
and determine the characteristics of X-ray/EUV spectra. 
For a neutron star with surface temperature 
$T\lo 10^6$ K and magnetic field $B\go 10^{12}$ G, 
various forms of hydrogen can be present in the envelope, 
including atom, poly-molecules, and condensed metal. 
We study the physical properties of different hydrogen phases 
on the surface of a strongly magnetized neutron star for 
a wide range of field strength $B$ and surface temperature $T$. 
Depending on the values of $B$ and $T$,
the outer envelope can be either in a 
nondegenerate gaseous phase or in a degenerate metallic phase.
For $T\go 10^5$ K and moderately strong magnetic field, 
$B\lo 10^{13}$ G, the envelope is nondegenerate and the surface material 
gradually transforms into a degenerate Coulomb plasma as density 
increases. For higher field strength, $B>> 10^{13}$ G, there exists a
first-order phase transition from the nondegenerate gaseous phase to the 
condensed metallic phase. The column density of saturated vapor 
above the metallic hydrogen decreases rapidly as the 
magnetic field increases or/and temperature decreases. 
Thus the thermal radiation can directly emerge from the 
degenerate metallic hydrogen surface. 
The characteristics of surface X-ray/EUV emission for different
phases are discussed. A separate study concerning the possibility of
magnetic field induced nuclear fusion of hydrogen on the neutron star
surface is also presented.  

\bigskip
{\it Subject headings:} stars: neutron -- 
stars: atmospheres -- magnetic fields --
atomic processes -- equation of state
-- radiation mechanisms: thermal 

\bigskip
\bigskip
\vfil\eject
\centerline{\bf 1. INTRODUCTION}
\nobreak
\bigskip

It has long been realized that neutron stars should remain detectable
as soft X-ray sources for $\sim 10^5$ years after their birth 
(Chiu \& Salpeter 1964; Tsuruta 1964). The last 
thirty years have seen significant progress in our understanding 
of various physical processes responsible to neutron star cooling 
(see e.g., Pethick 1992; Umeda, Tsuruta \& Nomoto 1994;
Reisenegger 1995). 
The advent of imaging X-ray telescopes has now made it possible to 
observe isolated neutron stars directly by their surface radiation.
In particular, ROSAT has detected pulsed X-ray thermal 
emission from a number of radio pulsars (see Becker 1995
for a review). Several nearby pulsars have also been detected by 
EUVE (Edelstein \& Bowyer 1996).
On the other hand, old isolated neutron stars ($10^8-10^9$ of 
which are thought to exist in the Galaxy), heated through accretion 
from interstellar material, are also expected to be common sources
of soft X-ray/EUV emission (e.g, Treves \& Colpi 1991; 
Blaes \& Madau 1993. See Walter et al.~1996 for a possible detection
and Pavlov et al.~1996 for spectral interpretations).
It has been suggested that some of the unidentified sources detected
in the EUVE and ROASAT WFC all-sky surveys may be associated with such
old neutron stars (Shemi 1995). 
Overall, the detections of the surface emission from neutron stars
have the potential of constraining the 
nuclear equation of state, various heating/accretion
processes, magnetic field structure and surface chemical composition.
Future observations are likely to extend to surface temperatures
as low as $10^5$ K. Confrontation with theory requires detailed
understanding of the physical properties of the outer layers of
neutron stars, in the presence of intense magnetic fields 
($B\go 10^{12}$ G) and low temperatures.

For young radio pulsars that have 
not accreted much gas, 
one might expect the surfaces to consist mainly of iron-peak 
elements formed at the neutron star birth. However, once the neutron stars 
accrete material, or have gone through a phase of accretion, either from 
the interstellar medium or from a binary companion, 
a hydrogen envelope will form on the top of the surface unless it is 
completely burnt out.
It is this envelope that mediates the radiation from the neutron star
surface. While the strong magnetic field and/or rapid stellar spin
may prevent large-scale accretion, 
it should be noted that even with a low accretion rate 
of $10^{10}$ g~s$^{-1}$ (typical of accretion from the interstellar
medium) for one year, the accreted material will be more 
than enough to completely shield the iron surface of a neutron star. 
The lightest elements, H and He, are likely to be the 
most important chemical species in the envelope due to their 
predominance in the accreting gas and also due to quick 
separation of light and heavy elements in the gravitational field 
of the neutron star (e.g., the settling time of C in a $10^6$ K hydrogen 
photosphere is of order a second). 
If the present accretion rate is low, 
gravitational settling produces a pure H atmosphere. 
If the temperature on the neutron star surface is not 
too high, light atoms, molecules and metal grains may form. 
The purpose of this paper is to study the phase diagram and the 
physical properties of the hydrogen envelope for a wide range of 
magnetic field strength and surface temperature. 

A strong magnetic field can dramatically change the structure 
of atoms, molecules and condensed matter 
(see Ruderman 1974 for an early review, and Ruder et al.~1994 for a 
recent text. Heavier atoms are also discussed by 
Lieb et al.~1994). The atomic unit $B_o$ 
for the magnetic field strength and a dimensionless parameter $b$ are
$$b\equiv {B\over B_o};~~~~B_o={m_e^2e^3 c\over\hbar^3}
=2.351 \times 10^9~{\rm G}.
\eqno(1.1)$$
When $b>>1$, the cyclotron energy of the electron
$\hbar\omega_e =\hbar (eB/m_ec)=11.58 B_{12}~{\rm keV}$,
where $B_{12}=b/425.4$ is the magnetic field strength in units of
$10^{12}~{\rm G}$, is much larger than the typical Coulomb energy, 
thus the Coulomb forces act as a perturbation to the magnetic 
forces on the electrons, and at most temperatures
the electrons settle into the ground Landau level.
Because of the extreme confinement of electrons in the
transverse direction, the Coulomb force becomes
much more effective for binding electrons in the parallel direction.
The atom has a cigar-like structure. Moreover, it is possible for
these elongated atoms to form molecular chains by covalent bonding along
the field direction. In two recent papers (Lai, Salpeter \& Shapiro 1992;
Lai \& Salpeter 1996; hereafter referred as Paper I and Paper II), 
we have studied the electronic structure and energy levels 
of various forms of hydrogen in strong magnetic 
field, including atoms, poly-molecular chains H$_N$ and condensed metal. 
In contrast to Fe chains, which is unbound at zero-pressure 
(Jones 1985; Neuhauser, Koonin \& Langanke 1987), 
we found that for typical magnetic field 
strength, $B_{12}\go 1$, the infinite H chains (thus the metallic 
hydrogen) are bound relative to individual atoms, and the cohesive 
energy increases with the field strength. This gives rise to the 
possibility of condensation of metallic hydrogen for sufficiently 
low temperature and/or high magnetic field. 
We quantify this phase transition in \S 4 of this paper.

In this study, we shall focus on the magnetic field strength in the
range of $10^{11}\lo B\lo 10^{15}$ G so that $b>>1$ is well-satisfied.
While field strengths of order $10^{12}-10^{13}$ G are considered 
typical for most neutron stars (with the exception of 
old millisecond pulsars and neutron stars in low-mass X-ray binaries),
it should be noted that the only physical upper limit to the neutron star
magnetic field strength is the virial equilibrium value,
$\sim 10^{18}$ G. Indeed, it has been suggested that 
magnetic fields as strong as $10^{15}$ G may be easily generated by 
dynamo processes in nascent neutron stars (Thompson \& Duncan 1993),
and may be required for soft gamma-ray repeaters (Paczy\'nski 1992;
Duncan \& Thompson 1993). 

Following the pioneering study of non-magnetic neutron star atmospheres
by Romani (1987), recent atmosphere modeling has taken account 
of the transport of different photon modes through an
ionized medium in strong magnetic field 
(Pavlov et al.~1994, 1995; Zavlin et al.~1995). 
Neutral atoms have been studied in detail for zero-field atmospheres
(Rajagopal \& Romani 1996; Zavlin et al.~1996).
Magnetic atoms have also been included in some models (Miller 1992), 
although many problems related to 
the treatment of the bound states in a strong magnetic field still
remain. One outstanding issue concerns the non-trivial coupling 
between the center-of-mass motion and internal structure of the atom
(e.g., Avron, Herbst \& Simon 1978; Herold, Ruder \& Wunner 1981).
This has been considered in detail in our recent paper (Lai \& Salpeter 
1995; hereafter Paper III) and used to derive the generalized
Saha equation for ionization-recombination equilibrium. With this,
we are now in a good position to calculate reliable composition 
and construct the complete phase diagram of the magnetic hydrogen 
surface of a neutron star. 
The published atmosphere models
have shown that neutral hydrogen is important 
for zero-field, cool atmospheres ($T\sim 10^{5}$ K),
but we shall see that even at $B_{12}\sim 0.01$ its importance 
extends to higher temperatures. Moreover, we shall find that 
at high field strength polyatomic molecules (for $B_{12}\go 1$)
and even the condensed phase (for $B_{12}\go 100$) are important.

Our paper is organized as follows. 
In \S 2 we discuss the physics of various hydrogen bound states 
in a strong magnetic field. In \S 3 we study the physical properties
of the nondegenerate atmosphere, including the abundance of various
species. We consider the phase equilibrium of the metallic state in \S 4. 
Some astrophysical implications of our results are discussed in \S 5. 
Appendix B includes a study of the magnetic field induced pycnonuclear
reactions that occur on neutron star surfaces. 

Throughout the paper we shall use real physical units and 
atomic units (a.u.) interchangeably, whichever is more convenient. 
Recall that in atomic units, mass and length are expressed in units 
of the electron mass $m_e$ and the Bohr radius 
$a_o=0.529\times 10^{-8}$ cm, energy in units of 
$2~{\rm Rydberg}=e^2/a_o=2\times 13.6$ eV; field strength 
in units of $B_o$ (Eq.~[1.1]), temperature in units of 
$3.15\times 10^5$ K, and pressure in units of $e^2/a_o^4=
2.94\times 10^{14}$ dynes~cm$^{-2}$.

\bigskip
\bigskip
\centerline{\bf 2. HYDROGEN BOUND STATES IN STRONG MAGNETIC FIELD}
\nobreak
\bigskip

Here we briefly review the physics of various forms of hydrogen bound
states in a strong magnetic field. The detailed quantum mechanical
calculations are described in Paper I and Paper II, where extensive 
references can be found. Our main results are summarized in various
fitting formulae for the binding energies and in Table 1. 

Our discussions are based on 
nonrelativistic quantum mechanics, even for extremely high magnetic field, 
$B\go B_{\rm rel}=(\hbar c/e^2)^2 B_o= 4.414 \times 10^{13}~{\rm G}$, 
at which the transverse motion of the electron 
becomes relativistic. This nonrelativistic treatment of bound states
is valid because (i) the electron remains nonrelativistic in the 
$z$-direction (along the field axis) as long as the binding energy 
is much less than $m_ec^2$, and (ii) 
the shape of the Landau wavefunction in the relativistic theory 
is the same as in the nonrelativistic theory, as the cyclotron radius
$$\hat\rho=\left({\hbar c\over eB}\right)^{1/2}={1\over b^{1/2}}~({\rm a.u.})
=2.57\times 10^{-10}B_{12}^{-1/2}~({\rm cm}),$$
is independent of the particle mass. 
Thus our results should be reliable even for the highest field strength 
considered in this paper. 
In \S 2.3, we shall include a discussion of 
the density-induced relativistic effect when we 
consider the metallic hydrogen state.

\bigskip
\centerline{\bf 2.1 Atoms}
\nobreak
\medskip

In a superstrong magnetic field satisfying $b>>1$,
the spectra of the H atom can be specified by two quantum numbers
$(m,\nu)$, where $m=0,1,2,\cdots$ measures the mean separation
$\rho_m=(2m+1)^{1/2}\hat\rho$ of the electron 
and proton in the transverse direction (perpendicular to the field),
while $\nu$ is the number of nodes of the electron's $z$-wavefunction 
(along the field axis). The states with $\nu \neq 0$ resemble 
the zero-field hydrogen atom with small binding energy 
$|E_\nu|\simeq 1/(2\nu^2)$ 
and we shall mostly focus on the tightly-bound states with $\nu=0$.
For the ground state $(0,0)$, the sizes (in atomic units) of the 
atomic wavefunction perpendicular and parallel to the field are of 
order $L_\perp\sim \hat\rho=b^{-1/2}$ and $L_z\sim l^{-1}$, where 
$l\equiv \ln b$. The binding energy $|E({\rm H})|$ (or the ionization
energy $Q_1$) of the atom is given by
$$Q_1=|E({\rm H})|\simeq 0.16\,l^2~({\rm a.u.})
\simeq 161\,\left[{\ln(426B_{12})\over\ln 426}\right]^2~({\rm eV}).
\eqno(2.1)$$
The numerical factor $0.16$ in equation (2.1) is an approximate
value for $B_{12}\go 1$ (More accurate fitting
can be found in Paper II). Some
numerical values are given in Table 1.
The tightly-bound excited states $(m,0)$ have 
the transverse size $L_\perp\sim\rho_m=[(2m+1)/b]^{1/2}$.
For $2m+1<<b$, we have $L_z\sim 1/\ln(1/\rho_m)$, and 
$$E_m\simeq -0.16\, l_m^2=-0.16\,\left(\ln{b\over 2m+1}\right)^2
~({\rm a.u.})~~~~~~({\rm for}~~2m+1<<b).
\eqno(2.2a)$$
For $2m+1\go b$, we have $L_z\sim \rho_m^{1/2}$, and the energy levels
are approximated by
$$E_m\simeq -0.6\,\left({b\over 2m+1}\right)^{1/2}
~({\rm a.u.})~~~~~~({\rm for}~~2m+1\go b).
\eqno(2.2b)$$
Note that, unlike the field-free case, the excitation energy
$\Delta E_m=|E({\rm H})|-|E_m|$ is small compared with $|E({\rm H})|$.

The above results assume a fixed Coulomb potential produced by the proton
(i.e., infinite proton mass). The use of a reduced electron mass
$m_em_p/(m_e+m_p)$ only introduces a very small correction
to the energy (of order $m_e/m_p$). 
However, as mentioned in \S 1, 
the effect of the center-of-mass motion on the energy spectra 
is rather subtle in strong magnetic field. 
We will come back to this point in Sec.~3.2 
(see Paper III).

\bigskip
\centerline{\bf 2.2 Molecules}
\nobreak
\medskip

In a superstrong magnetic field, the mechanism of forming
molecules is quite different from the zero-field case
(Ruderman 1974; Papers I-II). The spins of the electrons of the atoms
are all aligned anti-parallel to the magnetic field,
and therefore two atoms in their ground states ($m=0$) 
do not easily bind together according to the exclusion principle. 
Instead, one H atom has to be excited to the $m=1$ state. The
two H atoms, one in the ground state ($m=0$), another
in the $m=1$ state then form the ground state of the H$_2$ molecule by
covalent bonding. Since the ``activation energy'' for exciting
an electron in the H atom from Landau orbital $m$ to $(m+1)$
is small (see Eq.~[2.2]), the resulting molecule is stable.
In this way, more atoms can be added to form a larger molecule, 
in contrast to the field-free case.

For a given magnetic field, as the number of H
atoms $N$ increases, the electrons occupy more and more Landau orbits
(with $m=0,~1,~2,\cdots,N-1$), but the length of the chain decreases
so that the volume per electron of the electron distribution is
of order (neglecting logarithmic factors) $(bN)^{-1}$. 
Beyond some critical number 
$N_s\sim (b/l^2)^{1/5}$, it becomes energetically more favorable for 
the electrons to settle into the inner 
Landau orbitals (with smaller $m$) with nodes in their longitudinal
wavefunctions (i.e., $\nu\neq 0$). For $N\go N_s$, 
the energy per atom in the H$_N$ molecule,
$|E({\rm H}_N)|/N$, asymptotes to a value $\sim b^{2/5}$, 
independent of $N$, and the volume per electron to $\sim b^{-6/5}$
(Paper I; see \S 2.3). 
For a typical magnetic field strength of interest here, the energy
saturation point is $N_s\sim 3-5$. 

The dimensions of the H$_2$ molecule
parallel and perpendicular to the magnetic field 
are comparable to those of the atom. 
The binding energies also approximately scale as $(\ln b)^2$, 
but they are numerically smaller than the ionization energy of H atom.
More precisely, the {\it dissociation energy\/} 
of H$_2$ can be fitted by 
$$Q_2^{(\infty)}\equiv 2|E(\rH)|-|E(\rH_2)|
=0.106\,\left[1+\tau\ln\left(b/b_{\rm crit}\right)\right]\,
l^2~({\rm a.u.}),~~~~~\tau\simeq 0.1\,l^{0.2},
\eqno(2.3)$$
with an accuracy of $\lo 5\%$ for $1\lo B_{12}\lo 1000$, 
where $b_{\rm crit}=1.80\times 10^4$ is to be defined in equation (3.8)
(The superscript ``$(\infty)$'' implies that the zero-point energy 
of the molecule is not included in $Q_2^{(\infty)}$; see \S 2.4). 
Thus $Q_2^{(\infty)}\simeq 46$ eV for $B_{12}=1$ and 
$Q_2^{(\infty)}\simeq 150$ eV for $B_{12}=10$ (cf.~Table 1). 
By contrast, the zero-field dissociation energy of H$_2$ is $4.75$ eV.

For the ground state of H$_2$, the molecular axis and the magnetic field 
axis coincide, and the two electrons occupy the $m=0$ and $m=1$ orbitals, 
i.e., $(m_1,m_2)=(0,1)$. The molecule can have different types of 
excitation levels (Paper II): 

(i) {\it Electronic excitations}.
The electrons occupy different orbitals other than $(m_1,m_2)=(0,1)$, 
giving rise to the electronic excitations. The energy difference 
between the excited state $(m_1,m_2)$ (with $\nu_1=\nu_2=0$) 
and the ground state $(0,1)$ is approximately proportional to $\ln b$, 
as in the case for atoms. 
Typically, only the single-excitation levels (those with $m_1=0$ and 
$m_2>1$) are bound relative to two atoms in the ground states. 
Another type of electronic excitation is formed by two electrons
in the $(m,\nu)=(0,0)$ and $(0,1)$ orbitals.  The dissociation energy of
this weakly-bound state is of order a Rydberg, and does not depend
sensitively on the magnetic field strength.

(ii) {\it Aligned vibrational excitations}.
These result from the vibration of the protons along the magnetic field axis.
For the electronic ground state, the energy quanta
of small-amplitude oscillations is approximately given by
$$\hbar\omega_\parallel\simeq 0.13\,(\ln b)^{5/2}\mu^{-1/2} ({\rm a.u.})
\simeq 0.12\,(\ln b)^{5/2}~({\rm eV}),
\eqno(2.4)$$
where $\mu=m_p/2m_e=918$ is the reduced mass of the two protons
in units of the electron 
mass. Thus $\hbar\omega_{\parallel}\simeq 10$ eV at $B_{12}=1$ 
and $\hbar\omega_{\parallel}\simeq 23$ eV at $B_{12}=10$, in contrast to 
the vibrational energy quanta $\hbar\omega_{\rm vib}\simeq 0.52$ eV for 
H$_2$ molecule in zero magnetic field. 

(iii) {\it Transverse vibrational excitations}.
The molecular axis can deviate from the magnetic field
direction, precesses and vibrates around the field line.
Such an oscillation is the high-field analogy of
the usual molecular rotation. The important
difference is that here this ``rotation'' is constrained
around the magnetic field line. For the ground electronic state,
the excitation energy quanta $\hbar\omega_{\perp 0}$ is given by
$$\hbar\omega_{\perp 0}\simeq 0.125\,b^{1/2}(\ln b)\mu^{-1/2} ({\rm a.u.})
\simeq 0.11\,b^{1/2}(\ln b)~({\rm eV}),
\eqno(2.5)$$
where the subscript ``0'' indicates that 
we are at the moment neglecting the magnetic forces on the protons
which, in the absence of the Coulomb forces, lead to the cyclotron 
motion of the protons (see \S 2.4). 
Thus $\hbar\omega_{\perp 0}\simeq 14$ eV at $B_{12}=1$ and 
$\hbar\omega_{\perp 0}=65$ eV at $B_{12}=10$.

It is important to note that in strong magnetic field,
the electronic and vibrational (aligned and transverse) 
excitations are all comparable,  with
$\hbar\omega_{\perp 0}\go\hbar\omega_\parallel$. This is in contrast 
to the zero-field case, where one has 
$\Delta \eps_{\rm elec} \gg \hbar \omega_{\rm vib}
\gg \hbar \omega_{\rm rot}$.

\bigskip
\centerline{\bf 2.3 Infinite Chains and the Condensed Metallic State}
\nobreak
\medskip

For $N>>N_s$, the binding energy per atom in a H$_N$ molecule 
saturates, and the structure of the molecule is the same as
that of an infinite chain H$_{\infty}$. 
By placing a pile of parallel infinite chains close together
(with spacing of order $b^{-2/5}$ a.u.),
a three-dimensional condensed metal can be formed (e.g., body-centered
tetragonal lattice; Ruderman 1971). We offer our assessment 
on various calculations of the binding energy of magnetic Coulomb 
lattice. 

\bigskip
\centerline{\it 2.3.1 Uniform Electron Gas Model and its Extension}
\nobreak
\medskip

The binding energy of the magnetic metal at zero pressure can be 
estimated using the uniform electron gas model (e.g., Kadomtsev 1970).
Consider a Wigner-Seitz cell with radius $r_s$, volume $V_s=4\pi r_s^3/3$
(the mean number density of electron is then $n_e=1/V_s$). 
When the electron Fermi energy $p_F^2/(2m_e)=(n_eh^2c/2eB)^2/2m_e$ 
is less than the 
cyclotron energy $\hbar\omega_e$, or when the density satisfies
$$\eqalign{
n_e &\le n_{\rm Landau}={eB\over hc}{2\over h}(2m_e\hbar\omega_e)^{1/2}
={1\over \sqrt{2}\pi^2\hat\rho^3}
=0.0716\,b^{3/2}~({\rm a.u.});\cr
&~~~~r_{\rm s,Landau}=1.49\,b^{-1/2}~({\rm a.u.}),
}\eqno(2.6)$$
the electrons only occupy the ground Landau level. 
In the simple uniform electron gas model, 
the energy per cell of the metallic hydrogen can be written as
$$E_s(r_s)={3\pi^2\over 8b^2r_s^6}-{0.9\over r_s}~~({\rm a.u.}),
\eqno(2.7)$$
where the first term is the kinetic energy 
$E_k=(2\pi^4/3m_e)({\hbar^2c/eB})^2n_e^2$, and the second term is the
Coulomb energy. For the zero-pressure metal, we have $dE_s/dr_s=0$,
and the equilibrium $r_s$ and energy are then given by 
$$r_{s,0}\simeq 1.90\,b^{-2/5}~({\rm a.u.}),~~~~
E_{s,0}\simeq -0.395\,b^{2/5}~({\rm a.u.}).
\eqno(2.8)$$
The corresponding mass density is $\rho_{s,0}\simeq 560\,B_{12}^{6/5}$
g~cm$^{-3}$, which is much smaller than the Landau density,
defined in Eq.~(2.6), by a factor $0.48\,b^{-3/10}$. 
The pressure $P$ in a metal compressed to, say, twice the
zero-pressure density $\rho_{s,0}$, is a few times 
$r_{s,0}^{-3}|E_{s,0}|$, i.e., of order $P\sim 0.1\,b^{8/5}
\sim 10^3\,B_{12}^{8/5}$ (a.u.). Pressures in 
a neutron star surface layer of interest are much smaller than this
and we need to consider only $n_e<n_{\rm Landau}$.

We now discuss several corrections to the simple uniform electron gas 
model for metallic hydrogen. 

(i) {\it Relativistic effect}. As noted before, 
the use of non-relativistic equations for the transverse motion of 
the electrons is a good approximation even for 
$B\go B_{\rm rel}\simeq 137^2 B_o$. We can show
that the density-induced relativistic effect is also small.
The critical density $n_{\rm Landau}$
for onset of Landau excitation is still given by equation (2.6). 
The relativistic parameter is 
$x_e\equiv p_F/(m_ec)=(n_e/n_{\rm Landau})(2B/B_{\rm rel})^{1/2}$.
At zero-pressure density, using equation (2.8), we have 
$x_e\simeq 5.03\times 10^{-3} b^{1/5}$.
Thus near zero-pressure density, the relativistic effect is always 
negligible for the range of field strength of interest in this paper. 

(ii) {\it Coulomb exchange interaction}.
The exclusion principle for the electron results in an exchange
correction to the Coulomb energy. The Dirac
exchange energy (in atomic units) per electron is given by $E_{ex}=
-3F/(4br_s^3)$, where $F$ is a function of the ratio $n_e/n_{\rm Landau}$
(Fushiki, Gudmundsson \& Pethick 1989). 
The effect of this (negative) exchange interaction 
is to increase $r_{s,0}$ and $|E_{s,0}|$.

(iii) {\it Non-uniformity of the electron gas}. The Thomas-Fermi 
screening wavenumber $k_{TF}$ is given by (e.g., Ashcroft \& Mermin 1976)
$k_{TF}^2=4\pi e^2 D(E_F)$, where $D(E_F)={\partial n_e/\partial E_F}$ 
is the density of states per unit volume at the Fermi surface 
$E=E_F=p_F^2/(2m_e)$. Since $n_e=(2eB/h^2c)p_F$, 
we have $D(E_F)={n_e/2E_F}=({m_e/n_e})({2eB/h^2c})^2$, and
$$k_{TF}=\left({4\over 3\pi^2}\right)^{1/2}b\,r_s^{3/2}~({\rm a.u.}).
\eqno(2.9)$$
(More details on the electron screening in strong 
magnetic field, including the anisotropic effect, can be found in 
Horing 1969). The gas is uniform when the screening length $k_{TF}^{-1}$
is much longer than the particle spacing $r_s$, i.e., $k_{TF}r_s<<1$. 
For the zero-pressure metal, using equation (2.8), we have
$k_{TF}r_s \simeq 1.83$, independent of $B$. Thus even as 
$B\rightarrow \infty$, the electron non-uniformity must be considered 
for the zero-pressure metal. This effect can be studied using 
the Thomas-Fermi type statistical model, including the exchange
and the Weizs\"acker gradient corrections (see Fushiki et al.~1992 and
references therein).

\bigskip
\centerline{\it 2.3.2 Cohesive Energy of the Condensed State}
\nobreak
\medskip

Although the simple uniform electron gas model and its Thomas-Fermi
type extensions yield reasonable binding energy for the metallic 
state, their validity and accuracy in the strong field regime 
cannot be easily justified, and it is the difference 
between the energy of the metal and that of the
atom that determines whether the metal is bound.
One uncertainty concerns the lattice structure of
the metal, since the Madelung energy can be very different from
the Wigner-Seitz value (the second term in Eq.~[2.7])
for a non-cubic lattice. In principle, a three-dimensional
electronic band structure calculation is needed to
resolve this problem, as Jones (1985, 1986) has attempted 
for carbon and iron using density functional theory. However, 
the electron correlation has not been taken into
account, and, like the Thomas-Fermi type statistical models, 
the accuracy of density functional approximation
in a strong magnetic field is not yet known 
(for a review of density functional theory as applied to non-magnetic 
terrestrial solids, see, e.g., Callaway \& March 1984). 

We consider the self-consistent Hartree-Fock method 
(Neuhauser, Koonin \& Langanke 1987; Paper I) to be the most reliable 
approach to the problem. This has only been done for one-dimensional
chains. Our numerical results for the energy (per atom) of the
H$_\infty$ chain can be fitted (to within $2\%$ accuracy for
$B_{12}$ up to $\sim 10^3$) to a form similar to equation (2.8):
$$E_{\infty}=-0.76\,b^{0.37}~({\rm a.u.})=-194\,B_{12}^{0.37}~({\rm eV}).
\eqno(2.10)$$
Note that $|E_\infty|$ in equation (2.10) is larger than equation
(2.8) by a factor of $1.8$. The cohesive energy of the 1d chain 
(energy release in H$+$H$_{\infty}=$H$_{\infty+1}$) is given (to
within $\lo 10\%$ accuracy) by 
$$Q_{\infty}^{(\infty)}=
|E_{\infty}|-|E(\rH)| \simeq 0.76\,b^{0.37}-0.16\,(\ln b)^2
~({\rm a.u.})\eqno(2.11)$$
where the superscript ``$(\infty)$'' indicates 
that the zero-point energy correction
to the cohesive energy has yet to be included (\S 2.4). 

The energy difference $\Delta E_s=|E_{s,0}|-|E_\infty|$ between the
3d metal and the 1d chain must be positive 
and can be estimated by considering the interaction (mainly quadrupole
-quadrupole) between the chains. We have found (Appendix A) 
that this difference is probably between $0.4\%$ and $1\%$ 
of $|E_\infty|$. Therefore,
for hydrogen, {\it the binding of the three dimensional metal 
results mainly from the covalent bond along the magnetic field axis, 
not from the chain-chain interaction}.
The cohesive energy of the zero-pressure H metal is
$Q_s=Q_\infty+\Delta E_s$.

A note about the difference between iron and hydrogen is appropriate 
at this point. For Fe, it is has been found that at $B_{12}\sim 1-10$, 
the infinite chain is not bound relative to the atom
(Jones 1985; Neuhauser et al.~1987), contrary to what the original 
calculations (e.g., Flowers et al.~1977) indicate. 
Therefore, the chain-chain interaction must play a crucial role in
determining whether the three dimensional zero-pressure Fe metal is
bound or not. The main difference between Fe and H is that for the Fe
atom at $B_{12}\sim 1$, many electrons are populated 
in the $\nu\neq 1$ states, whereas for the H atom, as long as $b>>1$, 
the electron always settles down in the $\nu=0$
tightly-bound state. Therefore, the covalent bonding mechanism for forming 
molecules (cf.~\S 2.2) is not effective for Fe at $B_{12}\sim 1$. 
However, for sufficiently high B field, when $a_o/Z>>\sqrt{2Z+1}\hat\rho$, 
or $B_{12}>>100(Z/26)^3$, we expect the Fe chain to be bound in a
similar fashion as the H chain discussed here. 

\bigskip
\centerline{\bf 2.4 Zero-Point Energies and Relative Binding Energies}
\nobreak
\centerline{\bf of Different Forms of Hydrogen}
\nobreak
\medskip


For the H$_2$ molecule, the zero-point energy of the aligned vibration
is $\hbar\omega_\parallel/2$ (Eq.~[2.4]). Equation (2.5) for the 
zero-point energy of the transverse oscillation includes only the
contribution of the electronic restoring potential $\mu\omega_{\perp
0}^2R_\perp^2/2$. Since the magnetic forces on the protons also induce
a ``magnetic restoring potential'' $\mu\omega_p^2R_\perp^2/2$, 
the zero-point energy of the transverse oscillation is (Paper II)
$$\hbar\omega_\perp=\hbar (\omega_{\perp 0}^2+\omega_p^2)^{1/2}
-\hbar\omega_p,
\eqno(2.12)$$
where $\hbar\omega_p=\hbar eB/(m_pc)\simeq 6.3\,B_{12}$ eV is the
cyclotron energy of proton. Thus dissociation of H$_2$ taking into 
account the zero-point energies is 
$$Q_2= Q_2^{(\infty)}-\left({1\over 2}\hbar\omega_\parallel
+\hbar \omega_\perp\right).
\eqno(2.13)$$

Now consider the zero-point energy in the metallic hydrogen. 
Neglecting the magnetic force,
the zero-point energy $E_{zp}$ of a proton in the lattice is of order
$\hbar\Omega_p$, where $\Omega_p=(4\pi e^2n_e/m_p)^{1/2}$ 
is the ion plasma frequency. Thus
for the zero-pressure metal, we have 
$$E_{zp}\sim \hbar\Omega_p=0.040\,r_s^{-3/2}~({\rm a.u.})\simeq
0.015\,b^{3/5}~({\rm a.u.}).\eqno(2.14)$$
This is much smaller than the 
total binding energy (Eq.~[2.10]) unless $B_{12}\go 10^5$. 
This means that for the range of field strength considered in this paper, 
the zero-point amplitude is small compared to the lattice 
spacing. Thus quantum melting is not effective (Ceperley \& Alder 1980)
and the metal is a solid at zero temperature.
Accurate determination of $E_{zp}$ requires a detailed understanding 
of the lattice phonon spectra. At zero-field, Monte-Carlo simulations give 
$E_{zp}\simeq 3\hbar\Omega_p\eta/2$, with $\eta\simeq 0.5$ 
(Hansen \& Pollock 1973). For definiteness, we will adopt the same
value for $E_{zp}$ in a strong magnetic field.
Taking into account the magnetic effect on the proton, the corrected
cohesive energy of 1d chain is expected to be 
$$Q_\infty= Q_\infty^{(\infty)}-\left[{1\over 2}\hbar\Omega_p\eta
+\hbar \left({\omega_p^2\over 4}+\eta^2\Omega_p^2\right)^{1/2}
-{1\over 2}\hbar\omega_p\right],\eqno(2.15)$$
with $\eta\simeq 0.5$. 
As mentioned in \S 2.3, the relative binding energy between 3d
condensate and 1d chain is close to $0.4\%-1\%$ of $|E_\infty|$. 
We shall consider the condensed phase only for $B_{12}>10$,
where the ratio $Q_\infty/|E_\infty|$ ranges from $0.3$ to $0.7$ for 
$B_{12}=10-500$. For definiteness we shall express the cohesive energy
of the 3d condensate as 
$$Q_s=|E_{s,0}|-|E(\rH)|=Q_\infty+\Delta
E_s=(1+\alpha)Q_\infty,\eqno(2.16)$$
with $\alpha\simeq 0.01-0.02$. 

\midinsert
\vfil
\leftskip 0.25in    
\rightskip 0.25in   
\newdimen\digitwidth
\setbox0=\hbox{\rm0}
\digitwidth=\wd0
\catcode`?=\active
\def?{\kern\digitwidth}
$$\vbox{
\tabskip=2em plus3em minus1.5em
\halign to4.5in{
\hfil#\hfil&\hfil#\quad\hfil&\hfil#\hfil&\hfil#\hfil&\hfil#\hfil
&\hfil#\hfil\cr
\multispan6\hfil {\bf TABLE 1} \hfil\cr
\noalign{\smallskip}
\multispan6\hfil {\bf Relative Binding Energies (in eV) of H, H$_2$
and H$_\infty$}\hfil\cr
\noalign{\medskip \hrule \vskip1pt \hrule \medskip}
$B_{12}$ & $Q_1$ & $Q_2^{(\infty)}$ & $Q_2$& $Q_\infty^{(\infty)}$
& $Q_\infty$\cr
\noalign{\medskip \hrule \medskip}
0.1 &  76.4&   14 &   13 & --- & ---\cr
0.5 &  130 &   31 &   21 &   --- & ---\cr
  1 &  161 &   46 &   32 &   29 &   20\cr
  5 &  257 &  109 &   80 &   91 &   71\cr
  10 &  310 &  150 &  110 &  141 &  113\cr
  50 &  460 &  294 &  236 &  366 &  306\cr
  100 &  541 &  378 &  311 &  520 &  435\cr
  500 &  763 &  615 &  523 & 1157 & 964\cr
  1000 &  871 &  740 &  634 & 1630 & 1350\cr
\noalign{\medskip \hrule} 
}
}$$
\noindent NOTE: $B_{12}=B/(10^{12}\,{\rm G})$. $Q_1$ is
the ionization energy of H atom, $Q_2$ is the dissociation
energy of H$_2$, and $Q_\infty$ is the cohesive energy of 
1d H chain. The superscript ``$(\infty)$''
implies infinite proton mass. The given numerical results are
generally accurate to within $10\%$. 
Calculations of $Q_\infty$ are not reliable for $B_{12}\lo 1$.
\vskip 0.3in

\vfil
\endinsert

To summarize, we plot in Figure 1
the energy releases $Q_1,~Q_2$ and $Q_\infty$ for e+p=H, H+H=H$_2$, 
and H+H$_\infty=$H$_{\infty+1}$ respectively. Some numerical values
are given in Table 1. 
The zero-point energy corrections for $Q_2$ and
$Q_\infty$ have been included in the figure (if they are neglected, 
the curves are qualitatively similar, although 
the exact values of the energies are somewhat changed.)
Although $b>>1$ satisfies the nominal requirement for the
``strong field'' regime, a more realistic expansion
parameter for the stability of the condensed state over atoms
and molecules (cf.~Eqs.~[2.1],~[2.3],~[2.8] and [2.11]) is the ratio
$b^{0.4}/(\ln b)^2$. This ratio exceeds $0.3$, and 
increases rapidly with increasing field strength only for 
$b\go 10^4$. For $B_{12}\lo 10$, we see that $Q_1>Q_2>Q_\infty$
(cf.~Fig.~1), the condensed metallic phase may (or may not) exist in
the atmosphere. Fortunately, at the temperatures of interest 
in this paper we shall need to consider the condensed state 
only for $B_{12}\go 10$, where $Q_1>Q_\infty>Q_2$ 
(for $10\lo B_{12}\lo 100$) or $Q_\infty>Q_1>Q_2$ 
(for $B_{12}\go 100$). This will have
important consequences on the composition of the saturated vapor
above the condensed phase (\S 4).

\bigskip
\centerline{\bf 2.5 Surface Energy of Droplets and Large Molecules}
\nobreak
\medskip

For a single linear molecule H$_N$ we write $Q_N$ as 
the dissociation energy into H$_{N-1}+$H and denote the limit
for an infinite chain as $Q_\infty$. Numerical values and fitting
formulae for $Q_2$ and $Q_\infty$ are presented in \S\S 2.2-2.4. 
Results (without zero-point energy correction)
for $N=3,~4$ have also been given in Paper I. 
While accurate energies for larger molecules are needed in order to
establish the proper scaling of $Q_N$ as a function of $N$ and $B$,  
we note the following trend: 
The ratio $Q_2/Q_\infty$ (and to some extent $Q_3/Q_\infty$)
decreases appreciably with increasing $B$, but $Q_N/Q_\infty$
varies little with $B$ for large $N$. For a given $B$,
the ratio $Q_N/Q_\infty$ reaches a maximum near 
$N\sim N_s$ (the saturation point) and then approaches 
unity from above for $N>>N_s$.

For the phase equilibrium between the condensed metal and H$_N$
molecules in the vapor (\S 4) we will need the ``surface energy''
$S_N$, defined as the energy release in converting the 3d condensate
H$_{s,\infty}$ and a H$_N$ molecule into H$_{s,\infty+N}$. 
Clearly $S_1=Q_s=(\alpha+1)Q_\infty$ 
is the cohesive energy defined in \S\S 2.3-2.4. 
For linear H$_N$ molecule with energy (per atom) $E_N=E({\rm H}_N)$, 
we have 
$$S_N=N (E_N-E_s)=N\Delta E_s+N(E_N-E_\infty)
=N\alpha Q_\infty+fQ_\infty,\eqno(2.17)$$
where the first term on the right-hand side comes from cohesive
binding between chains, and the second term
is the ``end energy'' of a 1d chain. 
For $N\ge 2$, we can express $S_N$ in terms of $Q_N$'s via
$$S_N=NQ_s-\sum_{i=2}^NQ_i=
N\alpha Q_\infty+\left[Q_\infty+\sum_{i=2}^N(Q_\infty-Q_i)\right].
\eqno(2.18)$$
The dimensionless factor $f$ in Eq.~(2.17) is of order unity:
For $N=4$, it equals $0.23,~0.6,~1.1$ and $1.6$ for
$B_{12}=1,~10,~100$ and $500$, respectively.
The asymptotic values of $f$ for $N>>N_s$ should be close to these
values (since $Q_N\rightarrow Q_\infty$ for $N>>N_s$). 

Since $\alpha\sim 0.01$, representing the cohesion between 
infinite chains, is so small, the first term in equation (2.17)
is appreciable only when $N\go\alpha^{-1}\sim 100$. 
In that case, the molecular configuration which minimizes the surface
energy $S_N$ of the condensed metal is not a linear chain, but
some highly elongated ``cylindrical droplet'' with $N_\perp$
parallel chains each containing $N_\parallel$ atoms with 
$N_\perp N_\parallel=N$. For such a droplet, the ``end energy''
is of order $N_\perp fQ_\infty$. On the other hand there are
$\sim \sqrt{N_\perp}$ ``unpaired'' chains in such a droplet,
each gives an energy $N_\parallel \alpha Q_\infty$. Thus the total
surface energy is of order $\left[N_\perp f +\alpha
(N/N_\perp)\sqrt{N_\perp}\right]Q_\infty$. The minimum surface energy
$S_N$, for a fixed $N\go 2f\alpha^{-1}\equiv N_c$, is then obtained for
$N_\perp\simeq (N/N_c)^{2/3}$, $N_\parallel\simeq
N^{1/3}N_c^{2/3}$, and is of order 
$$S_N\simeq 3\,f\left({N\over N_c}\right)^{2/3}\!\!Q_\infty,
~~~~{\rm for}~N\go N_c\equiv {2f\over\alpha}\sim 200.\eqno(2.19)$$
Thus, although the optimal droplets are
highly elongated, the surface energy of the condensed metal still
grows as $(N/200)^{2/3}$ for $N\go 200$.

\bigskip
\centerline{\bf 2.6 Other Species: H$^-$ and H$_2^+$}
\nobreak
\medskip

For completeness, here we also give fitting formulae for the binding 
energies of the negative ion H$^-$ and the molecular ion H$_2^+$. The 
energy release in H+e=H$^-$ is 
$$Q({\rm H}^-)\simeq 0.014\,(\ln b)^2~({\rm a.u.})\eqno(2.20)$$
Thus the ionization energy of H$^-$ is $13$ eV for $B_{12}=1$ and 
$24$ eV for $B_{12}=10$, as compared to its value of $0.75$ eV for $B=0$. 
The energy release in H+p=H$_2^+$ is 
$$Q({\rm H}_2^+)\simeq (0.008+0.11\,\ln b)(\ln b)^2~({\rm a.u.})
\eqno(2.21)$$
Equations (2.20)-(2.21) are both accurate to within $20\%$. 

Because of the small binding energies, both H$^-$ and H$_2^+$ are 
likely to have negligible abundance in a typical neutron star atmosphere. 
However, the H$^-$ ion might, in principle, contribute appreciably to the 
atmospheric opacity. 

\bigskip
\bigskip
\centerline{\bf 3. THE SAHA EQUILIBRIUM AND WARM ATMOSPHERES}
\nobreak
\bigskip

We now consider the physical conditions and chemical equilibrium 
in neutron star atmospheres with photospheric temperature 
in the range $T_{\rm ph}\sim 10^5-10^{6.5}$ K and magnetic field
strength in the range $B_{12}\sim 0.1-20$. These conditions
are likely to be satisfied by most observable neutron stars. 
For such relatively low field strengths,
the atmosphere consists mainly of ionized hydrogen, 
H atoms and small H$_N$ molecules, and we can neglect the condensed 
metallic phase in the photosphere.  
Although the density scale height of the atmosphere is only 
$h\simeq k T_{\rm ph}/ m_p g \simeq 0.08\,T_{\rm ph,5}\,g_{14}^{-1}$
(cm), where $T_{\rm ph,5}=T_{\rm ph}/(10^5~{\rm K})$, $g$ is surface
gravitational acceleration, $g_{14}=g/(10^{14}\,{\rm cm~s}^{-2})$,
the atmosphere has significant optical depth. 
In \S 4 we shall consider more extreme situation 
($B_{12}>> 10$) when the non-degenerate atmosphere has negligible
optical depth and the condensed metallic phase becomes
important.

Although the surface layers we consider can lie somewhat below
the photosphere, we shall restrict to densities smaller than 
the zero-pressure condensate density (cf.~\S 2.3)
$$\rho_{s,0}\simeq 560\,B_{12}^{6/5}~({\rm g~cm}^{-3}).\eqno(3.1)$$
Thus the density is always below the critical ``Landau density''
(cf. Eq.~[2.6]) 
$$\rho_{\rm Landau}=7.08\times 10^3\,\mu_e B_{12}^{3/2}
~({\rm g~cm}^{-3}),\eqno(3.2)$$ 
where $\mu_e$ is the mean molecular weight per electron 
($\mu_e=1$ for ionized hydrogen). For our ranges of $B$ and 
$T_{\rm ph}$ we also have the
inequality $T<<\hbar\omega_e/k=1.34\times 10^8B_{12}$ K, so that 
all electrons settle in the ground Landau state. The electron Fermi
temperature $T_F=E_F/k$ is then given by 
$$T_F= 2.67 B_{12}^{-2}\left({\rho \over\mu_e}\right)^2~({\rm K}),
\eqno(3.3)$$
where $\rho$ is the mass density in g~cm$^{-3}$.
At the density of the condensate (Eq.~[3.1]), we would have
$T_F\simeq 8.4\times 10^5B_{12}^{2/5}$ K; but 
for $\rho$ much smaller than $\rho_{s,0}$, we have
$T_{\rm ph}>>T_F$, and the electrons 
are nondegenerate.\footnote{$^1$} 
{Equation (3.3) should be compared with the Fermi temperature
in field-free case, $T_F(0)=3.0 \times 10^5(\rho/\mu_e)^{2/3}~
({\rm K})$. Clearly, high magnetic field lifts the degeneracy of
electrons even at relatively high density.} 

\bigskip
\centerline{\bf 3.1 The Photospheric Density and Pressure}
\nobreak
\medskip

In a star with an interior energy source (such as an isolated cooling
neutron star), the photon flux $F=\sigma_{\rm sb}T_{\rm eff}^4$ is
constant to very far below the photosphere ($\sigma_{\rm sb}$ is the
Stefan-Boltzmann constant). With an accretion energy 
source, this constancy holds only above the stopping layer for the 
infalling matter. However, the optical depth $\tau_{\rm stop}$
in this layer is much larger than unity for a warm 
atmosphere\footnote{$^2$}
{The column density $y_{\rm stop}$ of the stopping layer typically
corresponds to a few to tens of Thompson depths (depending on the field
strength), i.e., $y_{\rm stop}\sim 10/\kappa_{\rm es}$, where 
$\kappa_{\rm es}$ is the electron scattering opacity (cf.~Nelson
et.~al.~1993). Thus for an absorption-dominated atmosphere,
$\tau_{\rm stop}\sim 10\kappa_{\rm ff}/\kappa_{\rm es}>>1$.},
and we shall consider only optical depth 
$\tau$ less than $\tau_{\rm stop}$, where 
$\tau=\int\kappa\,dy$ with $\kappa$
the Rosseland mean opacity and $y=\int\rho\,dz$ the column density.
With the flux $F=\sigma_{\rm sb}T_{\rm eff}^4$ assumed constant, a
simple approximation for the temperature profile $T(\tau)$ is 
$$T^4(\tau)\simeq {3\over 4}T_{\rm eff}^4\left({2\over 3}+\tau\right),
\eqno(3.4)$$
and the photospheric temperature $T_{\rm ph}=T_{\rm eff}$ (with 
$\tau_{\rm ph}$ chosen as $2/3$). With the gravitational 
acceleration $g$ in the atmosphere constant, 
hydrostatic equilibrium gives for the pressure $P=gy$
and, rewriting $\tau=\bar\kappa y$, we have
$$P(\tau)={g\over\bar\kappa}\tau;~~~~
P_{\rm ph}\simeq {2g\over 3\bar\kappa}.\eqno(3.5)$$
Note that Eq.~(3.5) holds irrespective of the equation of state, 
even when there is a phase transition at some pressure (see \S 4).
For the warm atmospheres considered in this section, the ideal gas
law applies and the photospheric density is given by 
$\rho_{\rm ph}\simeq P_{\rm ph}m_p/(kT_{\rm ph})$.

For the temperature and density relevant to the atmosphere, 
free-free absorption dominates over electron scattering. 
In the zero-field case, assuming Kramer's opacity 
$\kappa(0)=\kappa_0 \rho T^{-3.5}$, with $\kappa_0 \simeq 7.4\times
10^{22}$ (cgs), we would obtain the  photosphere density-temperature
relation $\rho_{\rm ph}(0)
\simeq 7\times 10^{-3} g_{14}^{1/2}\,T_{\rm ph,5}^{5/4}
~{\rm g\,cm}^{-3}$. In a strong magnetic field,
the radiative opacity becomes anisotropic and depends on polarization
(Canuto et al.~1971; Lodenquai et al.~1974; Pavlov et al.~1994). 
For photons with polarization vector perpendicular to the magnetic field
(the ``extraordinary mode'')
the free-free absorption and electron scattering opacities
are reduced below their zero-field values by a factor of
$(\omega/\omega_e)^2$, while for photons polarized along the magnetic
field (the ``ordinary mode''), the opacities are not affected.
Silant'ev and Yakovlev (1980) have calculated the appropriate average
Rosseland mean free-free opacity. In the field and temperature regime
of interest, an approximate fitting formula is
$$\kappa(B)\simeq 400\,\beta \left({kT \over \hbar \omega_e}\right)^2
\kappa(0),\eqno(3.6)$$
with $\beta\simeq 1$.
The resulting photosphere density and pressure are given by
$$\eqalign{
\rho_{\rm ph} (B) &\simeq 0.5\,\beta^{-1/2}\,g_{14}^{1/2}
\,T_{\rm ph,5}^{1/4}\,B_{12}~({\rm g\,cm}^{-3}),\cr
P_{\rm ph}(B) &\simeq 4\times 10^{12}
\,\beta^{-1/2}\,g_{14}^{1/2}\,T_{\rm ph,5}^{5/4}\,B_{12}
~({\rm dyne\,cm}^{-2}).\cr
}\eqno(3.7)$$
In Figure 2, this photosphere ($\tau\simeq 2/3$) condition is shown
in a temperature-density diagram. For $\beta\sim 1/400$, equation
(3.7) also approximately characterizes the physical conditions of the
deeper layer where the extraordinary photons are emitted. This
``$\tau_\perp\simeq 2/3$'' line is also shown in Figure 2. 
Solving Eqs.~(3.4)-(3.5) (more precisely, $dP/d\tau=g/\kappa$),
we can obtain the temperature profile
of the atmosphere as a function of density. Some of such $T-\rho$ 
profiles for different values of $T_{\rm eff}=T_{\rm ph}$ are depicted
in Figure 2. Clearly, $T\rightarrow T_{\rm ph}/2^{1/4}=0.84T_{\rm ph}$
as $\tau\rightarrow 0$, while $T\propto\tau^{1/4}\propto\rho^{4/9}$
as $\tau\rightarrow\infty$.

Other sources of opacity such as bound-free and bound-bound absorptions
will increase the opacity and reduce the
photosphere density, but the above estimates
define the general range of the physical parameters
in the atmosphere if $T_{\rm eff}$ is large enough for the neutral H 
abundance to be small.
This rough estimates agree
reasonably well with more detailed 
calculations of Pavlov et al.~(1995). 

\bigskip
\centerline{\bf 3.2. Ionization Saha Equilibrium}
\nobreak
\medskip

We now consider the ionization-recombination equilibrium of the
H atom. Previous treatments of this problem (e.g., Khersonskii 1987;
Miller 1992) 
have assumed that the H atom can move across the magnetic field freely.  
This is generally not valid for the strong field regime of interest here. 
A free electron confined to the
ground Landau level, the usual case for $b>>1$,
does not move perpendicular to the magnetic field. Such motion
is necessarily accompanied by Landau excitations.
When the electron combines with a proton,
the mobility of the neutral atom across the field
depends on the ratio of the atomic excitation energy ($\sim \ln b$)
to the Landau excitation energies $\hbar\omega_p=\hbar eB/(m_pc)$ 
for the proton. It is convenient to define a critical field strength 
$B_{\rm crit}$ via
$$b_{\rm crit}\equiv {m_p\over m_e}\ln b_{\rm crit}=1.80\times 10^4;~~~
B_{\rm crit}=b_{\rm crit}B_o=4.23\times 10^{13}~{\rm G}.
\eqno(3.8)$$
Thus for $B\go B_{\rm crit}$, the deviation from the free center-of-mass
motion of the atom is significant (Paper III).

In a strong magnetic field, the center-of-mass motion of the atom is 
specified by a pseudomomentum ${\bf K}$, whose component along the 
field axis is simply the usual liner momentum, while the 
perpendicular component ${\bf K}_\perp$ measures the mean transverse
separation of the electron and proton (Avron, Herbst \& Simon 1978;
Herold, Ruder \& Wunner 1981; Paper III and references therein). 
For a H atom in the ground state, the total 
energy (in atomic units) is approximately given by (Paper III)
$${\cal E}_0(K_z,K_{\perp}) \simeq {K_z^2\over 2M}-|E(\rH)|
+{K_c^2\over 2M_\perp}\ln\left(1+{K_\perp^2\over K_c^2}\right),
\eqno(3.9)$$
where we have defined 
$$M_\perp \simeq M\left(1+{\xi\,b\over Ml}\right)
\simeq M\left(1+{\xi\,b\over b_{\rm crit}}\right),~~~
K_c^2\simeq 0.64\,\xi\,b\left(1+{Ml\over \xi\,b}\right)^2,
\eqno(3.10)$$
and $M=m_p+m_e\simeq m_p/m_e$ (a.u.), $\xi\simeq 2.8$, 
$|E(\rH)|\simeq 0.16\,l^2$ (a.u.)
(cf.~Eqs.~[2.1]-[2.2]). Note that for $K_\perp<<K_c$, the dependence
of ${\cal E}_0$ on $K_\perp$ becomes $K_\perp^2/(2M_\perp)$. Thus 
$M_\perp$ represents the effective mass for the transverse motion
across the magnetic field. Potekhin (1994) has given more 
accurate numerical results for a few selected field strength,
but the approximation in equation (3.9) is adequate for our purpose. 

Let the proton (free or bound) number density in the gas be $n_g$. 
The partition function (in atomic units) for a H atom in a volume 
$V_g=4\pi r_g^3/3=1/n_g$ is then 
$$Z(\rH)\simeq V_g\left({MT\over 2\pi}\right)^{1/2}{M_\perp'T\over 2\pi}
\exp\left({|E(\rH)|\over T}\right)z(\rH),
\eqno(3.11)$$
where 
$$M_\perp'=M_\perp\left(1-{2M_\perp T\over K_c^2}\right)^{-1}
\eqno(3.12)$$
for $T<< K_c^2/(2M_\perp)$, and $M_\perp'/M_\perp$ approaches a
constant less than unity when $T$ becomes comparable to
$K_c^2/(2M_\perp)$. The partition function $Z(\rH)$ has the same form as
the zero-field expression 
except for the factor $M_\perp'/M$. In equation (3.11),
$z(\rH)=z_m(\rH)z_\nu(\rH)$, where
$z_m(\rH)$ is the internal partition function associated with
the $m>0$ excited states:
$$z_m(\rH)\simeq \left(1+e^{-b/MT}\right)
\sum_{m=0}^{m_{\rm max}}{M_{\perp m}'\over M_\perp'}
\exp\left[-{1\over T}\left(0.16\,l^2+E_m+m{b\over M}\right)\right],
\eqno(3.13)$$
where $M_{\perp m}'$ is of the same order of magnitude as $M_\perp'$, 
the factor $\left(1+e^{-b/MT}\right)$ comes from the proton spin 
effects, and $m_{\rm max}$ is set by the condition $L_\perp\lo r_g$
or $L_z\lo r_g$ (cf.~\S2.1).
 The internal partition function $z_\nu$ associated with the
$\nu>0$ states is close to unity (see Paper III for detail). 

The partition functions for the ionized states (free electron and proton) 
can be easily obtained. 
With the atomic bound-state partition function given above, 
we can obtain the generalized Saha equation for the 
ionization-recombination equilibrium in the presence of strong
magnetic field. In the density and temperature regimes of interest, 
with $T\lo K_c^2/(2M_\perp)\simeq 0.32\,l\,(1+Ml/\xi b)$, we have
$${n(\rH)\over n_p n_e}\simeq \left({b\over 2\pi}\right)^{-2}\!\!
M_\perp'\left({T\over 2\pi}\right)^{1/2}\!\!\tanh\left({b\over
2MT}\right)\tanh\left({b\over 2T}\right)
\exp\left({Q_1\over T}\right)z(\rH),
\eqno(3.14)$$
where $n({\rm H})$, $n_p$ and $n_e$ are the number
densities of the different species.

\bigskip
\centerline{\bf 3.3 Dissociation Equilibrium of Molecules}
\nobreak
\medskip

Accurate treatment of the dissociation equilibrium for H$_N$ molecules
is complicated. However, since the molecular excitation
energies are comparable to the excitations in H atom (\S 2.2), 
we expect that when 
$T\lo K_c^2/(2M_\perp)\sim l\,(1+Ml/\xi b)$, 
we can similarly use an effective mass description 
for the motion of the molecule across the magnetic field. 
As an estimate, we assume that the effective mass
of H$_N$ molecule is $M_{\perp}({\rm H}_N)\sim NM(1+\xi b/Ml)$. 
We similarly introduce the correction factor $M_\perp'/M_\perp$ given
in equation (3.12).
The Saha equation for the equilibrium process
H$_N+$H$=$H$_{N+1}+Q_{N+1}$ is then given by 
$${n({\rm H}_{N+1}) \over n({\rm H}_N) n({\rm H})}
\simeq \left({N+1\over N}\right)^{3/2}\!\!
\left({MT\over 2\pi}\right)^{-3/2}\!\left({M\over M_\perp'}\right)
\!\exp\left({Q_{N+1}\over T}\right){z({\rm H}_{N+1})\over 
z({\rm H}_N) z({\rm H})},
\eqno(3.15)$$
where $z({\rm H}_{N+1})$, $z({\rm H}_N)$, $z({\rm H})$ 
are the internal partition functions
of H$_{N+1}$, H$_N$, H respectively. 
To estimate the internal partition functions of the molecules, 
one need to include various molecular excitation levels as discussed 
in \S 2.2. But presumably the ratio of these internal partition
functions is of order unity (this approximation is
adopted in the calculations presented in \S 3.5), 
since the molecular excitation level 
spacing is not necessarily smaller than the electronic excitation (\S
2.2). 

\bigskip
\centerline{\bf 3.4 Non-ideal Gas Effect}
\nobreak
\medskip

So far we have assumed the gas to be sufficiently dilute to be treated
as an ideal gas. When the gas density increases, 
the interaction between particles
becomes important. The effect of the atom-atom interaction potential 
$U_{12}(r)$ on the ionization equilibrium is 
to modify the chemical potential of the atomic gas by the amount
$$\Delta\mu(\rH)=n(\rH)kT\,\int\!d^3r(1-e^{U_{12}/kT}),\eqno(3.16)$$
(e.g., Landau \& Lifshitz 1980).
For $r$ much larger than the atomic length $L_z\sim 1/l$ of the
elongated atom, 
the potential $U_{12}$ is due to the quadrupole-quadrupole interaction 
$U_{12}\sim Q^2(3-30\cos^2\theta+35\cos^4\theta)/r^5$,
where $\theta$ is the angle between the vector 
${\bf r}$ and the $z$-axis, and $Q\sim eL_z^2$ is the quadrupole 
moment of the atom. Since the integration over
the solid angle $\int\!d\Omega U_{12}=0$ at large $r$, the
contribution to $\Delta\mu(\rH)$ from  large $r$
is negligible. The main effect is then the ``excluded volume
effect'': Let $v_a$ be the volume of the (highly
non-spherical) electron distribution for one atom, so that we can set 
$U_{12}\rightarrow\infty$ due to quantum mechanical repulsion when two
atoms overlap. We then have 
$\Delta\mu(\rH)\sim kTn(\rH)v_a$. The atom-atom interaction introduces a
factor $\exp(-\Delta\mu/kT)\sim\exp(-nv_a)$ 
to the right-hand-side of the Saha equation (3.14). 
Therefore when the mean density 
in the gas becomes comparable to the internal density of
the atom, the atom is mostly ionized --- so called {\it pressure
ionization}. One can similarly consider 
the interaction between a charged particle (electron or proton)
and the atom, given by the quadrupole-monopole potential
$U_{12}\sim eQ(3\cos^2\theta-1)/r^3$.
Again since $\int d\Omega U_{12}=0$, the long-range 
term is negligible. We thus obtain a similar factor of order
$\sim \exp(-nv_a)$ correction to the Saha equation.

The ``atomic volume'' $v_a$ is of order $\pi L_\perp^2L_z$, 
where $L_\perp\sim b^{-1/2}$ is the radius perpendicular to the
z-axis and $L_z\sim 1/l$ the length in the z-direction.
The ratio of this volume to the volume per electron in the
condensed state, $4\pi r_{s,0}^3/3$ with $r_{s,0}$ given in equation
(2.8), is then $\sim 0.1\,b^{1/5}/\ln b$, which increases slowly 
with increasing $b$ and is of order $0.1$ for $B_{12}\sim 1-10^3$.
The ``excluded volume'' of the atom may be larger due to its
elongated shape, but $v_a\lo L_z^3\sim
4\pi r_{s,0}^3/3$. Thus we have
$\Delta \mu(\rH)\sim kT\,(\rho/\rho_{s,0})$,
where $\rho_{s,0}$ given by Eq.~(3.1). 
The corresponding pressure-ionization factor in the Saha equations is
$\exp(-\Delta\mu/kT)$. 
 

\bigskip
\centerline{\bf 3.5 Results}
\nobreak
\medskip

To get a qualitative overview of the relative abundances of different
forms of hydrogen, we have determined the
H atom half-ionization and H$_2$ half-dissociation curves in the
$T-\rho$ diagram. In Figure 2, we show the results for
$B_{12}=1$ and $B_{12}=10$; we also plot the electron degeneracy line
($T=T_F$) and the typical photosphere conditions (\S 3.1). 
The half-ionization line is determined by setting $n_e/n_g=n_p/n_g
=n({\rm H})/n_g=1/2$, 
while the half-dissociation line is obtained with $n({\rm H})/n_g=1/2$
and $n({\rm H}_2)/n_g=1/4$ (so that H and H$_2$ have equal mass
fractions). Note that the gas density is simply $\rho=m_p n_g$, and
the pressure ionization factor discussed in \S 3.4 is not 
included in the calculations.  
To the left of the half-ionization line (labeled
``e$+$p$=$H'') the atmosphere is mostly ionized, while
to the right of the half-dissociation line (labeled
``H$+$H$=$H$_2$''), most of the material is in the molecular states,
H$_2$, H$_3$, etc. Between the ionization line and the H$_2$
dissociation line, the dominant species is H atom. 
From Eqs.~(3.7) and (3.14), we can obtain the 
photospheric ``ionization temperature'', $T_{\rm ph,ir}$,
where the ionization-recombination equilibrium gives $50\%$
H and $50\%$ e$+$p in the photosphere ($\tau=2/3$). 
For $B_{12}\sim 1-50$, an approximate fitting formula is
$$\log_{10}T_{\rm ph,ir}\simeq 5.4+0.3\log_{10}B_{12}.
\eqno(3.17)$$
If the actual photospheric temperature $T_{\rm ph}$ is greater 
than $T_{\rm ph,ir}$, then there is little neutral H 
near and above the photosphere ($\tau\lo 1$),
and the neutral abundance increases slowly with depth. 
If $T_{\rm ph}$ is only slightly smaller than $T_{\rm ph,ir}$,
then neutral H dominates in layers near the photosphere 
($\tau\sim 1$), but (e$+$p) dominates at higher levels ($\tau<<1$), 
because $T$ is almost constant for $\tau\lo 1$ while $\rho$
decreases to zero as $\tau\rightarrow 0$ (cf.~Fig.~2). 
For $B_{12}=1$, the abundances of molecular species are negligible
unless the photospheric temperature drops well below $10^5~{\rm K}$. 
For $B_{12}=10$, there exists a large amount of H$_2$ in the
photosphere when $T_{\rm ph,5} \lo 2$. 

The ``neutrality fraction'' $n(\rH)/n_g$ as a function of temperature
at a fixed density $\rho=n_g m_p=0.01$ g/cm$^3$
and $0.1$ g/cm$^3$ is shown in Fig.~3.
Results for a wide range of field strengths
($B_{12}=0.01,\,0.1,\,1,\,10$) are given and compared with 
the zero-field limit. In general, the neutral fraction is
not a monotonic function of $B$, because it 
is determined by two opposite effects [cf.~Eq.~(3.14)]:
the ionization energy $Q_1$ increases with increasing $B$, 
this tends to increase $n(\rH)/n_g$; on the other hand, 
the phase space of free electron and proton, proportional to 
$b^2/\tanh(b/2MT)$, also increases with increasing
$B$, and this tends to decrease $n(\rH)/n_g$. 
It is also easy to understand from Eq.~(3.14) that the neutral
fraction is in general a non-monotonic function of $T$ at a given
field strength. As seen from Fig.~3, even a relatively small field 
increases the neutrality appreciably above the field-free
value, partly because $Q_1$ is larger, but also because the internal
partition function $z_m(\rH)$ is larger. Thus it may well 
be that a $B=10^9-10^{10}$ G atmosphere (which characterizes 
millisecond pulsars) is very different from a field-free
atmosphere (Rajagopal \& Romani 1996; Zavlin et al.~1996).

Figure 4 shows the fraction (by
number) of various hydrogen gas species 
as a function of $T_{\rm ph}$ at a point with 
$P=0.5P_{\rm ph}$ (cf.~Eq.~[3.7] with $\beta=1$ and $g_{14}=1$),
$\tau=1/3$ and $T=(3/4)^{1/4}T_{\rm ph}=0.93T_{\rm ph}$. 
Three magnetic field strengths ($B_{12}=1,~10,~100$) are considered. 
The total number density $n_{\rm tot}$ of free particles in the gas, i.e., 
$n_{\rm tot}=n_e+n_p+n({\rm H})+n({\rm H}_2)+\cdots$, 
is given by $n_{\rm tot}=P/kT=1.6\times 10^{23}B_{12}T_{\rm ph,5}^{1/4}$
(cm$^{-3}$).
Neglecting the trace ion species such as H$^-$ and H$_2^+$ (and the
condensed phase),
the abundances satisfy the condition $2X_p+X({\rm H})
+X({\rm H}_2)+X({\rm H}_3)+\cdots=1$, where 
$X_p=n_p/n_{\rm tot}$ and $X(\rH_N)=n(\rH_N)/n_{\rm tot}$.
For determining $X_p,~X(\rH),~X(\rH_2)$, the values of $Q_1,~Q_2$
given in Table 1 are used, while for determining the fractions of
larger molecules, the following approximate {\it ansatz} is adopted in
our calculation (cf.~\S 2.5): For H$_3$ and H$_4$, the ratio 
$Q_N^{(\infty)}/Q_\infty^{(\infty)}$ given in Paper I 
is assumed, and the zero-point energy correction to $Q_N$ 
is introduced by setting
$Q_N=(Q_N^{(\infty)}/Q_\infty^{(\infty)})Q_\infty$ 
with $Q_\infty$ given in Table 1; For $N>4$, where no numerical
result is currently available, we simply set $Q_N=Q_\infty$. 
Clearly, the uncertainty is largest for 
intermediate-sized molecules ($N\sim N_s$), but we have checked 
that moderate variations (less than $30\%$) of these 
intermediate $Q_N$'s do not change our results significantly. 
The approximation should become better for $N>>N_s$
as $Q_N$ asymptotes to $Q_\infty$. 
For $B_{12}=1,~10$, we see that the atmosphere is dominated by
atoms and relatively small molecules, and the total gas density 
$\rho=m_p n_g=m_p[n_p+n(\rH)+2n(\rH_2)+3n(\rH_3)+\cdots]$
remains much less than the condensation density $\rho_{s,0}$. 
For the extreme field strength $B_{12}=100$, which we include
here for illustrative purpose, increasingly large molecules
dominate the atmosphere as $T_{\rm ph}$ decreases, and the
gas density approaches $\rho_{s,0}$. This is 
indicative of a phase transition which will be discussed in \S 4. 


For layers deeper than the photosphere, i.e., for optical depth
$\tau$ increasing to large values, the ionization fraction 
changes slowly since the temperature-density track in the atmosphere,
$\rho\propto T^{9/4}$ (for $\tau>>1$), is close to the track of
constant (e$+$p)/H ratio. Indeed, from Fig.~2 we see that 
the ionization fraction varies by only a factor of a few 
below the photosphere. 
The zero-pressure density $\rho_{s,0}$ of the condensed metal is given
by Eq.~(3.1) and shown by the heavy vertical lines in Figure 2. 
Note that the half-ionization lines and half-dissociation lines are
meaningful only in the density regime to the left of the 
condensation density lines and the degeneracy lines. 
When $\tau$ is sufficiently large that $\rho$ approaches $\rho_{s,0}$,
non-ideal gas effects and/or condensation set in. With 
$T\propto\rho^{4/9}$, the temperature at $\rho\sim\rho_{s,0}$
is of order $T\sim 20\,T_{\rm ph}$. For a warm atmosphere this exceeds
the critical temperature for phase transition, 
$T_{\rm crit}\sim 10^5B_{12}^{0.4}$ K (see \S 4). Thus there
is no distinct condensed phase, although the atoms would overlap at 
$\rho\go\rho_{s,0}$ and be pressure-ionized. The electrons form a
uniform fluid, but stay in the ground Landau level until $\rho$
exceeds $\rho_{\rm Landau}$ (the dot-dashed vertical lines in Fig.~2).


\bigskip
\bigskip
\centerline{\bf 4. ULTRAHIGH FIELDS AND COOL ATMOSPHERES:}
\nobreak
\centerline{\bf THE CONDENSED PHASE}
\nobreak
\bigskip

As discussed in \S 2, the cohesive energy $Q_s$ and
surface energy $S_N$ of the condensed metallic hydrogen increase
more rapidly with increasing field strength than the ionization energy
$Q_1$ of H atom and the dissociation energy $Q_N$ of small
H$_N$ molecules (cf.~Fig.~1).
We shall see below (cf.~Eqs.~[4.5]-[4.7]) that $Q_s=S_1$,
$S_N$ and $(Q_s+Q_1)/2$ determine the partial saturation
vapor densities of H, H$_N$ and (p$+$e) respectively.  
For $B_{12}>>10$, the cohesive energy is greater than 
$Q_1$ and $Q_2$, and is close to 
$|E_{s,0}|\simeq 8\,B_{12}^{0.4}$ a.u. In this case there 
may exist a critical temperature $T_{\rm crit}$, below
which there is a first order phase transition between the 
condensed metallic hydrogen\footnote{$^3$}
{The condensed metal may be solid or
liquid. Using the zero-field criterion as an estimate, 
the melting temperature is determined by
$\Gamma=e^2/(r_skT)=18.7B_{12}^{2/5}/T_5\simeq 180$, where
we have used the zero-pressure value for $r_s$
given by Eq.~(2.8). Thus for $B_{12}\lo 287\,T_5^{5/2}$ the
zero-pressure metal is likely to be a liquid.}
and the gaseous vapor. We explore the possibility 
of this phase separation and the partial saturation vapor pressure
of various constituents in this section. 

\bigskip
\centerline{\bf 4.1 Phase Equilibrium Conditions}
\nobreak
\medskip

Consider the equilibrium between the condensed metallic phase (labeled
by the subscript ``s'') and the non-degenerate gaseous phase (labeled
by ``g''). The electron number density in the metallic phase is 
$n_s=1/V_s=3/(4\pi r_s^3)$, where $r_s$ is the mean proton spacing
(the radius of Wigner-Seitz cell). The gaseous phase consists of a
mixture of free electrons, protons, bound atoms and molecules, with
the total baryon number density $n_g=1/V_g=n_p+n(\rH)+2n(\rH_2)+\cdots$.
We neglect the small concentration of H$^-$, H$_2^+$ and other
molecular ions in the gas, thus we have $n_e=n_p$. 
Phase equilibrium requires the temperature, pressure and the
chemical potentials of different species to satisfy the conditions: 
$$\eqalign{
&T_s=T_g=T,\cr
&P_s=P_g=[2n_p+n(\rH)+n(\rH_2)+n(\rH_3)+\cdots]kT=P,\cr
&\mu_s=\mu(\rH)=\mu_e+\mu_p={1\over 2}\mu(\rH_2)={1\over 3}\mu(\rH_3)
=\cdots\cr
}\eqno(4.1)$$

First consider the equilibrium between the condensed metal and 
H atoms in the gaseous phase. Near the zero-pressure metal 
density, the electron Fermi temperature
is $T_{F,0}\simeq 8.4\times 10^5B_{12}^{2/5}$ K (cf.~Eqs.~[3.1], [3.3]).
Thus at a given temperature, the metal becomes more
degenerate as $B$ increases. Let the energy per Wigner-Seitz
cell be $E_s(r_s)$. The pressure and 
chemical potential of the condensed phase are given by 
$$P_s=-{1\over 4\pi r_s^2}{dE_s\over dr_s},~~~~
\mu_s=E_s(r_s)+P_sV_s\simeq E_{s,0}+P_sV_{s,0},\eqno(4.2)$$
where the subscript ``$0$'' indicates the zero-pressure values, 
and we have assumed that the vapor pressure is 
sufficiently small so that the deviation from the zero-pressure state
of the metal is small, i.e., $\delta\equiv |(r_s-r_{s,0})/r_{s,0}|<<1$. 
This is justified when the saturation vapor pressure $P_{\rm sat}$
is much less than the critical pressure $P_{\rm crit}$ for phase
separation, or when $T<<T_{\rm crit}$.
The finite temperature correction $\Delta\mu_s$ to the 
chemical potential of the condensed phase is given by
\footnote{$^4$}
{For $kT<<\hbar\omega_e$ and $E_F<<\hbar\omega_e$ so that electrons
are all in the ground Landau level, we have
$n_s=(2eB/h^2c)\int_0^\infty\!dp_z\{\exp[(p_z^2/2m_e-\mu)/kT]+1\}^{-1}
\simeq (2eB/h^2c)(2m_e\mu)^{1/2}[1-(\pi kT)^2/(24\mu^2)]$,
which gives $\mu\simeq E_F[1+(\pi kT)^2/(12E_F^2)]$.
Note that this correction has the same form as that for a 
non-magnetic electron gas, but with opposite sign.}
$${\Delta\mu_s(T)\over kT}
\simeq {\pi^2\over 12}{T\over T_F}
\simeq 0.10B_{12}^{-2/5}T_5,
~~~~\Delta\mu_s(T)\simeq 0.85B_{12}^{-2/5}T_5^2~
({\rm eV}),\eqno(4.3)$$
where the Fermi energy $E_F=kT_F=9\pi^2/(8b^2r_s^6)\simeq
0.236\,b^{2/5}$ (a.u.), and for the density we have used the
zero-pressure value. The partition function of H atoms is given by 
equation (3.11). Using $n(\rH)=\exp[\mu(\rH)/T]Z(\rH)/V_g$ and the
equilibrium condition $\mu(\rH)=\mu_s$, we obtain the number density
of H atoms in the saturated vapor: 
$$n(\rH)\simeq \left({MT\over 2\pi}\right)^{3/2}\!
\left({M_\perp'\over M}\right)z(\rH)
\exp\left[{-Q_s+PV_{s,0}+\Delta\mu_s(T)
-\Delta\mu(\rH)\over T}\right],\eqno(4.4)$$
where the cohesive energy $Q_s=|E_{s,0}|-|E(\rH)|$ is given by 
equation (2.16). Since $PV_{s,0}$ is typically much smaller than the
cohesive energy $Q_s$, and for low vapor density we can neglect the
non-ideal gas correction $\Delta\mu(\rH)$, equation (4.4) reduces to
an explicit expression for $n(\rH)$: 
$$n(\rH)\simeq \left({MT\over 2\pi}\right)^{3/2}
\!\left({M_\perp'\over M}\right)z(\rH)\exp\left({-Q_s+\Delta\mu_s
\over T}\right).\eqno(4.5)$$

The densities of the other species in the 
saturated vapor can also be obtained. 
Equation (4.5) together with the Saha equation 
for e$+$p$=$H$+Q_1$ (Eq.~[3.14]) yields
$$n_p\simeq {bM^{1/4}T^{1/2}\over (2\pi)^{3/2}}
\left[\tanh\left({b\over 2MT}\right)\right]^{-1/2}\!\!\!
\exp\left({-Q_1-Q_s+\Delta\mu_s\over 2T}\right).\eqno(4.6)$$
The partition function of H$_N$ molecules can be approximated by 
$$Z(\rH_N)=V_g\left({NMT\over 2\pi}\right)^{3/2}\!
\left({M_\perp'\over M}\right)z(\rH_N)\exp\left(-{NE_N\over T}
\right),$$
(cf.~\S 3.3, and recall that $E_N$ is the energy
per atom in a H$_N$ molecule).
The equilibrium condition $N\mu_s=\mu_N$ 
for the process $\rH_{s,\infty}+\rH=\rH_{s,\infty+N}$ (with energy
release $S_N=NE_N-NE_s$,
where $S_N$ is the surface energy discussed in \S 2.5.) 
would then give
$n(\rH_N)=\exp(\mu_N/T)Z(\rH_N)/V_g$. For large molecular chains
($N>>N_s$) or droplets ($N\go N_c$; cf.~\S2.5), however,
a correction $\Delta\mu_N$ to the ``internal'' chemical potential
due to finite temperature need to be included ($\Delta\mu_N$
is to be distinguished from the non-ideal gas correction discussed
in \S3.4). For large 3d droplets, this correction is identical
to that given in Eq.~(4.3), i.e., $\Delta\mu_N=N\Delta\mu_s$.
For linear chains with $N>>N_s$, 
we expect that $\Delta\mu_N/N$ is still close to
$\Delta\mu_s$ since electrons behave like a Fermi gas
in the molecular chain as they do in 3d condensate,
although it is not clearly which is greater.\footnote{$^5$}
{Since the mean electron density in 3d condensate is higher than 
that in 1d chain by about $0.2\%$ (cf.~Appendix A), we may expect that
$\Delta\mu_N/N$ is larger than $\Delta\mu_s$ by $\sim 0.4\%$ according
to Eq.~(4.3).}
We therefore write $\Delta\mu_N=\zeta N\Delta\mu_s$, with $\zeta\simeq
0$ for $N\lo N_s$ and $\zeta\simeq 1$ for $N>>N_s$.
With this correction, the saturated density of H$_N$ is given by
$$n(\rH_N)\simeq N^{3/2}\left({MT\over2\pi}\right)^{3/2}
\!\left({M_\perp'\over M}\right)
z(\rH_N)\exp\left[{-S_N+N(1-\zeta)\Delta\mu_s\over T}\right].
\eqno(4.7)$$
For concreteness, we shall set $\zeta=0$ for $N=1-5$ and 
$\zeta=1$ for $N>5$ in our calculations below. This is adequate
for our purpose since $N(1-\zeta)\Delta\mu_s$ is typically 
much less than $S_N$ for small molecules. More
detailed (and unknown) prescription for $\zeta$ as a function of
$N$ (as long as $\zeta$ satisfies the asymptotic behavior discussed
above) would have negligible effect on our results presented below. 


\bigskip
\centerline{\bf 4.2 Critical Temperature for Phase Separation}
\nobreak
\medskip

The critical temperature $T_{\rm crit}$, below which phase 
separation between condensed metal and gaseous vapor occurs, is
determined
by the condition $n_s=n_g=n_p+n(\rH)+2n(\rH_2)+3n(\rH_3)+\cdots$. 
In equation (2.17) for the surface energy, the factor $f$
is of order unity, and approaches a constant for $N>>N_s$. 
Substituting Eq.~(2.17) with $f\sim$ constant
into Eq.~(4.7) with the approximation $z(\rH_N)\sim z(\rH)$, 
we obtain
$$n_g\simeq n_p+\left({MT\over2\pi}\right)^{3/2}
\!\!\!\left(\!{M_\perp'\over M}\!\right)
z(\rH)\exp\!\left(\!-{fQ_\infty\over
T}\!\right)\!\sum_N\!N^{5/2}\exp\!\left\{\!-{N[\alpha Q_\infty
-(1-\zeta)\Delta\mu_s]\over
T}\right\}.\eqno(4.8)$$
Clearly, for the sum to converge, we require $\alpha
Q_\infty>(1-\zeta)\Delta\mu_s$. With $\alpha Q_\infty=\Delta E_s\simeq 
0.01 |E_\infty|\simeq 2\,B_{12}^{0.37}$ eV, this reduces
to $T_5\lo 1.5\,B_{12}^{0.4}/(1-\zeta)^{1/2}$, which 
is easily satisfied for $T<T_{\rm crit}$ (see below) and
$\zeta\simeq 1$. Thus we neglect $(1-\zeta)\Delta\mu_s$
in comparison to $\alpha Q_\infty$.
The sum in Eq.~(4.8) therefore reduces to a finite number of order
$\bar N^{5/2}\,[\exp\left(\alpha Q_\infty/T\right)-1]^{-1}
\sim \bar N^{5/2}$ (recall that $fQ_\infty>>\alpha Q_\infty$),
where $\bar N\sim 2.5T/(\alpha Q_\infty)$ specifies 
the term that contributes most to the sum.
The critical temperature for phase transition is then given by 
$$T_{\rm crit}\simeq {fQ_\infty\over\ln\Lambda}\sim 0.1fQ_\infty,~~~~
\Lambda\equiv n_s^{-1}\!
\left({MT\over2\pi}\right)^{3/2}
\!\!\!\left({M_\perp'\over M}\right)z(\rH)\bar N^{5/2},
\eqno(4.9)$$
where we have neglected $n_p$, and in estimating $\ln\Lambda$ we 
have used $n_s=\rho_s/m_p\simeq 50\,B_{12}^{6/5}$ (a.u.)
(cf.~Eq.~[3.1]) and $T_5\sim 2\,B_{12}^{0.4}\sim T_{\rm crit}$
(which gives $\bar N\sim 20$). 
Thus for $f\simeq 1$, we find $T_{\rm crit}
\simeq 10^5,~5\times 10^5$ and $10^6$ K 
for $B_{12}=10,~100$ and $500$ respectively (see also Fig.~5).
The corresponding critical pressure is of order 
$P_{\rm crit}\simeq (n_g/\langle N\rangle)kT_{\rm crit}
\sim 0.1 n_sQ_\infty/\langle N\rangle$, where 
$\langle N\rangle$ is the typical size of molecules in the vapor. 


\bigskip
\centerline{\bf 4.3 Saturated Vapor of Condensed Metallic
Hydrogen}
\nobreak
\medskip

Figure 5(a)-(c) depicts the partial saturation vapor densities of 
different species in equilibrium with metallic hydrogen as 
a function of temperature for $B_{12}=10,~100$ and $500$.
These densities are calculated using Eqs.~(4.5)-(4.7),
where the surface energy $S_N=(N\alpha+f)Q_\infty$ 
is obtained from $Q_N$ and $Q_\infty$
via Eq.~(2.18), and we adopt the same {\it ansatz} for $Q_N$ as described
in \S3.5, i.e., we use the numerical results for
$Q_1,\,Q_2,\,Q_3,\,Q_4$ and $Q_\infty$ (the zero-point energy
corrections to $Q_3$ and $Q_4$ are incorporated via 
$Q_N/Q_\infty=Q_N^{(\infty)}/Q_\infty^{(\infty)}$), and we set 
$Q_N=Q_4$ for $N>4$. This approximation for $Q_N$ corresponds to
$f=0.23,\,0.6,\,1.1,\,1.7$ for $N\ge 4$ at $B_{12}=1,\,10,\,100,\,500$
respectively. In calculating the total baryon density $n_g$
in the vapor, we include linear molecular chains H$_N$
with $N$ up to $N_{max}=200$ [cf.~Eq.~(2.19)]
and neglect ``droplets'' in the vapor. 
Choosing a different $N_{max}>>1$ would only change
the total vapor density $n_g$ slightly for $T\lo T_{\rm crit}$. 
The solid vertical line in each panel of Fig.~5 indicates
the temperature at which $n_g$ equals $n_s$. This defines
the critical temperature $T_{\rm crit}$ for phase transition, 
and it is well approximated by $0.1\,Q_\infty/k$.
Thus $T_{\rm crit}\simeq 9\times 10^4,\,5\times 10^5$ and $1.4\times
10^6$ K for $B_{12}=10,~100$ and $500$ respectively.
Note that for illustrative purpose, we have considered in Fig.~5(a)
relatively low field strength, $B_{12}=10$, where the condensed
phase is more uncertain. This uncertainty for the low-field case is
reflected by the fact that $n(\rH_2)>n(\rH)$ in the saturated vapor,
so that larger molecules have even greater abundances. Thus there may
not be a distinct condensed phase at all, and even if phase separation
exists, the condensation temperature is below $10^5$ K. 
In such relatively low-$B$ regime, the outer layer of the neutron star
is characterized by {\it gradual\/} transformation from
nondegenerate gas to degenerate plasma as the pressure (or column
density) increases (see \S 3). For $B_{12}>>10$ [cf.~Fig.~5(b)-(c)],
on the other hand, the vapor density becomes much less than
the condensation density as the temperature decreases and/or the
magnetic field increases, thus phase separation is 
inevitable.  

The most abundant species in saturated vapor at 
$T\sim T_{\rm crit}/2$ (for example)
depend on the field strength, with smaller fields favoring
poly-atomic molecules since $f$ in Eq.~(2.17) is smaller for smaller
$B$. The ratio $n(\rH_2)/n(\rH)$ depends exponentially on 
$(S_1-S_2)/T=(Q_2-Q_s)/T\simeq (Q_2-Q_\infty)/T$.
From Figure 1, we see that $Q_\infty-Q_2>0$ for
for $B_{12}\go 10$, and increases with increasing $B$. 
Therefore, when $B_{12}>>10$,
there are few H$_2$ molecules compared with H atoms in the
saturated vapor. The ionization ratio $n_p/n(\rH)$ in the
vapor depends exponentially on $(Q_s-Q_1)/(2T)$.
From Figure 1, we see that $Q_\infty-Q_1>0$ when $B_{12}\go 200$, 
in which case the vapor mostly 
consists of ionized hydrogen (cf.~Fig.~5(c)). 
For $10\lo B_{12}\lo 200$,
$n(\rH)$ is greather than $n_p$ and $n(\rH_2)$, 
although the abundances of larger molecules
are also appreciable.


The column density $y_{\rm sat}$ above the surface of the condensed
phase is related to the saturation vapor pressure $P_{\rm sat}$ 
by $y_{\rm sat}=P_{\rm sat}/g$ and is plotted against $T$
in Figure 6 for $B_{12}=10,~100,~500$ (all for $g_{14}=1$). 
Note that $T$ is the temperature at the gas/metal phase boundary.
The density of the vapor is typically 
$\rho\sim m_pgy/(kT)\sim 10\,g_{14}T_5^{-1}y$ (g~cm$^{-3}$).
Using equation (3.6) for the magnetic free-free opacity, 
we obtain the optical depth of the vapor above the metallic
hydrogen, $\tau_{\rm ff}\sim 400\beta g_{14}T_5^{-5/2}B_{12}^{-2}y^2$.
The threshold vapor column density $y_{\rm th}$, 
below which the vapor is optically thin, is then given by 
$$y_{\rm th}\sim 0.05\,\beta^{-1/2}g_{14}^{-1/2}B_{12}\,T_5^{5/4}~
({\rm g\,cm}^{-3}).
\eqno(4.10)$$
This threshold value $y_{\rm th}$ is also plotted against $T$ 
in Fig.~6. The intersect of the $y_{\rm sat}(T)$ and 
$y_{\rm th}(T)$ curves defines a ``threshold photospheric 
temperature'' $T_{\rm ph,th}$ for the optical depth 
above the the surface of the condensate (if any):
For $T_{\rm ph}>T_{\rm ph,th}$, the photosphere
is purely in the gaseous phase, as described in \S 3. 
For $T_{\rm ph}<T_{\rm ph,th}$ the vapor above the condensed metal
is optically thin, and there is a sharp change in density at the 
gas/metal interface; The ``photosphere'' is located inside 
the condensed phase. 
For $B_{12}=500$, the saturated vapor is dominated by ionized 
hydrogen, hence the value for $T_{\rm ph,th}$ 
should be fairly accurate. For $B_{12}=10$ and $100$, 
however, the vapor is dominated by atoms or molecules, 
free-free opacity is only an estimate and thus the values
of $T_{\rm ph,th}$ given in Fig.~6 are more approximate. 
In any case, we see that $T_{\rm ph,th}$ is only slightly less than
the condensation temperature $T_{\rm crit}$. 

In the case of $T_{\rm ph}<T_{\rm ph,th}$, the pressure at the
gas/metal interface $P_{\rm sat}$ is less than $P_{\rm ph}$. 
Some distance into the condensed phase, with pressure increasing 
from $P_{\rm sat}$ to $P_{\rm ph}$ (which is still much less than 
$P_{\rm crit}$ for the condensed phase), the liquid is
still optically thin and $T$ stays close to $T_{\rm ph}$. Further still
into the metal, as $P$ increases toward $P_{\rm crit}$ from the
``liquid photosphere'', the temperature increases appreciably while
the density increases only slowly. Whether this region is 
convective or radiative depends on the opacity law
in the condensed metal, which we have not yet investigated. 

\bigskip
\bigskip
\centerline{\bf 5. DISCUSSIONS}
\nobreak
\bigskip

Recent works on the equation of state of neutron star surface in 
strong magnetic field have focused on the outer crust consisting of 
iron-like elements (e.g., Fushiki et al.~1989;
Abrahams \& Shapiro 1991). Heat transport through this crustal region 
can be affected by the strong magnetic field 
(e.g., Hernquist 1985). However, for a neutron star 
covered by hydrogen atmospheric layer,
the outer crust of the solid lattice is not the directly observable 
region of the star. Recent theoretical modeling of neutron star
atmosphere aims at interpreting ROSAT's soft X-ray spectra from
several radio pulsars, and most studies have focused on ionized
atmospheres (Pavlov et al.~1995; Zavlin et al.~1995).  


Our present study indicates that for sufficiently low temperature
($T\lo 10^6$ K) and/or high magnetic field ($B\go 10^{12}$ G), 
hydrogen atoms or molecules can have large abundance
in the photosphere (see Fig.~2). This relatively large neutral 
abundance (even at relatively high temperature)
comes about mainly for two reasons: (i) The binding energies of the
bound states are greatly increased by the magnetic field; 
(ii) The photospheric density is much larger because the surface
gravity is strong and also because the opacity of one of the two
photon modes is significantly reduced by the magnetic field.
Therefore one could in principle 
expect some atomic or molecular line features in the soft X-ray or UV 
spectra. In particular, the Lyman ionization edge ($160$ eV
at $10^{12}$ G and $310$ eV at $10^{13}$ G) is expected to be
prominent by the following consideration:
The free-free and bound-free cross-sections for a photon in the 
extraordinary mode (with the photon electric field perpendicular 
to the magnetic field) are approximately given by 
(Gnedin, Pavlov \& Tsygan 1974; Ventura et al.~1992; 
Potekhin \& Pavlov 1993):
$$\sigma_{{\rm ff}\perp}\simeq 1.7\times 10^3\rho T_5^{-1/2}
\alpha^2a_o^2\omega^{-1}b^{-2},~~~~~
\sigma_{{\rm bf}\perp}\simeq 4\pi\alpha a_o^2\left({Q_1\over\omega}
\right)^{3/2}\!\!b^{-1},
$$
where $\alpha=1/137$ is the fine structure constant, 
$a_o$ is the Bohr radius, $\rho$ is the density in g$\,$cm$^{-3}$, 
$\omega$ is the photon energy in the atomic units, $Q_1$ is the ionization 
potential, and $b$ is the dimensionless field strength
defined in Eq.~(1.1). 
Near the absorption edge, the ratio of the free-free and
bound-free opacities is
$${\kappa_{{\rm ff}\perp}\over\kappa_{{\rm bf}\perp}}
\sim 10^{-4}{\rho\over T_5^{1/2}B_{12}X(\rH)}.
$$
Thus even for relatively small neutral H abundance $X(\rH)$,
the discontinuity of the total opacity at the Lyman edge is 
pronounced.\footnote{$^6$}
{Since the extraordinary mode has smaller opacity, most of the 
X-ray flux will come out in this mode. For photons in the ordinary 
mode, we have $\sigma_{{\rm ff}\parallel}\simeq 1.65\times 10^3
\rho T_5^{-1/2}\alpha^2a_o^2\omega^{-3}$ (the zero-field value), and 
$\sigma_{{\rm bf}\parallel}\simeq 10^2\pi\alpha
a_o^2(Q_1/\omega)^{5/2}(\ln b)^{-2}$, which give
$\kappa_{{\rm ff}\parallel}/\kappa_{{\rm bf}\parallel} 
\sim 10^{-2}\rho T_5^{-1/2}$ for $B_{12}\sim 1$. 
Thus the absorption edge for the
ordinary mode is less pronounced.}
Clearly, the position and the strength of the absorption 
edge can provide a useful diagnostic of the neutron star surfaces.

For even stronger magnetic field ($B$ larger than a few times
$10^{13}$ G), we have shown that the degenerate matter 
can extend all the way to the outer edge of the star with little 
gaseous atmosphere above it, if the effective surface temperature
$T_{\rm eff}$ is less than a threshold value $T_{\rm ph,th}$
(cf.~Fig.~6). 
In this case, the thermal radiation
arises from just below the metallic hydrogen surface. Using equation
(3.1) as an estimate for the condensation density (near zero presure),
we find that the electron plasma frequency $\Omega_e$ is given by 
$\hbar\Omega_e\simeq 0.66\,B_{12}^{3/5}$ keV, much larger than the
typical photon energy $kT\simeq 10\,T_5$ eV. Thus photons cannot be
easily excited thermally inside the metallic surface. 
As a result, the surface emission is reduced from 
the black-body emission with the same temperature, and
the spectrum of the surface radiation is likely to 
deviate quite strongly from a black-body spectrum (e.g.,
Itoh 1975; Brinkmann 1980). We have not investigated the opacity
structure of the hydrogen condensate and cannot say whether the
emission mimics highly diluted blackbody radiation with color
temperature much larger than the effective surface temperature.


Finally, we note that the possibility that the outmost layer of a
neutron star consists of degenerate iron has been considered
previously, based on the notion that in a strong magnetic field the 
cohesive energy of Fe solid can be 
as large as tens of keV (Ruderman 1974). 
While it is certainly correct that a strong magnetic field 
can significantly enhance the cohesive energy, 
it has been shown (Neuhauser et al.~1987) that the earlier calculation
(Flowers et al.~1977) greatly overestimated the cohesive energy of Fe
solid (cf.~\S2.3.2). Thus the metallic iron surface scenario was
replaced by a non-degenerate atmosphere model. Our results in this
paper partially resurrects the ``degenerate surface picture'': 
hydrogen (not iron) can condensate at sufficiently high B field and
low temperature. 

\bigskip
\medskip

This work has been supported in part by
NASA Grant NAGW-2394 and NAG 5-2756 to Caltech, and 
by NSF Grant AST 91-19475, NASA Grant NAGW-666 and NAG 5-3097
to Cornell University. D.L. also acknowledges the
support of a Richard C. Tolman Research Fellowship in theoretical
astrophysics at Caltech. 

\bigskip
\bigskip
\bigskip
\centerline{\bf APPENDIX A. ESTIMATE THE RELATIVE BINDING ENERGY}
\nobreak
\centerline{\bf BETWEEN 1D CHAIN AND 3D CONDENSATE}
\nobreak
\bigskip

In the absence of self-consistent electronic structure calculation
for the 3d condensed metal, we estimate the fractional relative
binding energy, $\Delta
E_s/|E_\infty|=(E_s-E_\infty)/|E_\infty|$, using the following three
methods:

(i) {\it Method 1}:
If we approximate the 1d chain by an uniform cylinder (radius
$R$, proton spacing $a$), then the energy per atom (subcylinder)
in the chain can be written as
$$E_\infty={2\pi^2\over 3b^2R^4a^2}-{1\over a}
\left[\ln{2a\over R}-\left(\gamma-{3\over 4}\right)\right],
\eqno(A1)$$
where $\gamma=0.57721566$ (see Paper I).
Minimizing $E_\infty$ with respect to $R$ and $a$
gives
$$R=1.694\,b^{-2/5},~~~~a/R=1.885,~~~~E_\infty=-0.3913\,b^{2/5}.\eqno(A2)$$
Comparing with equation (2.8) for the energy of 3d metal in the
Wigner-Seitz approximation, we find $\Delta E_s/|E_\infty|\simeq -1\%$. 

(ii) {\it Method 2}: Imagine that the 3d condensate is formed
by placing a pile of parallel chains in contact with each other. 
The chain is assumed to have the property given by equation (A2). 
For a body-centered tetragonal lattice, the relative 
position of nearest neighboring protons is $x_0=2R$ perpendicular to
the field and $z_0=a/2$ parallel to the field. We calculate
the electrostatic interaction energy $E_{12}$ between 
a pair of nearest neighboring subcylinders using 
a Monte-Carlo integration method (Interaction 
between subcylinders of larger separations can be calculated
using a quadrupole-quadrupole formula, but this interaction
is small compared to the nearest neighbor interaction). 
This gives $E_{12}=-3.7\times 10^{-4}b^{2/5}$. Since each 
proton has eight nearest neighbors, we have 
$\Delta E_s=4E_{12}$, and thus $\Delta E_s/|E_\infty|\simeq -0.4\%$. 

If we choose $x_0<2R$ (which would involve overlap of electron
clouds), the chain-chain interaction is stronger. 
For $x_0=\sqrt{2}R$ (the smallest value possible), we
find $\Delta E_s/|E_\infty|\simeq -8\%$. We consider this as
an upper limit to the true relative binding energy. 

(ii) {\it Method 3}: Similar to (ii), we form 3d solid by 
placing chains together into a body-centered tetragonal lattice,
with $x_0=2R$ and $z_0=a/2$. But instead of equation (A2), we determine
$R$ and $a$ by minimizing the total energy of the solid,
$E_s=E_\infty+\Delta E_s$, where $E_\infty$ is given by equation (A1),
and the numerical result for $\Delta E_s=4E_{12}$ can be fitted
by the following analytic formula (with accuracy $\lo 1\%$ for
$1.6\le a/R\le 2.2$):
$$\Delta E_s=-4.0\times 10^{-4}{1\over R}\left[1+23
\left({a\over R}-1.4\right)^2\right].
\eqno(A3)$$
Minimizing $E_s(R,a)$, we obtain:
$$R=1.650\,b^{-2/5},~~~~a/R=2.036,~~~~E_s=-0.3932\,b^{2/5}.\eqno(A4)$$
Comparing with equation (A2) for a single chain, we find 
$\Delta E_s/|E_\infty|\simeq -0.5\%$. 

\bigskip
\bigskip
\bigskip
\centerline{\bf APPENDIX B. PYCNONUCLEAR REACTIONS INDUCED BY}
\nobreak
\centerline{\bf STRONG MAGNETIC FIELD}
\nobreak
\bigskip

In this appendix, we consider the interesting possibility of ``cold
fusion'' of hydrogen bound states induced by strong magnetic field on the 
neutron star surfaces. The magnetized H molecules and condensed metal 
are highly compressed, i.e., the internal electron density within 
these bound states is large. This results
in strong screening to the ion-ion Coulomb 
potential, and, as we will show, dramatic increase 
in the fusion rate between the nuclei within the bound states. 

Nuclear reaction in bound molecular systems has been studied 
in the context of muon-catalysed fusion, where a massive 
muon replaces the electron in a diatomic molecule of hydrogen isotope,
enhancing the binding and producing large
cold fusion rates 
(e.g., Zel'dolvich \& Gershtein 1961).
In astrophysics, nuclear reactions induced by high matter
density (`pycnonuclear reactions') have also been studied
(e.g., Salpeter \& Van Horn 1969; Schramm \& Koonin 1990).
The magnetic field induced fusion is similar to 
the density-induced pycnonuclear reaction except that 
the large density increase comes from the formation of bound states 
in strong magnetic field.

We consider the enhancement of the fusion rate by magnetic field 
in both molecules and condensed metals. 

\bigskip
\centerline{\bf B.1 Internal Fusion of H$_2$}
\nobreak
\medskip

Consider a diatomic molecule composed of two isotopic nuclei 
of hydrogen. The fusion rate $\Lambda$ is proportional to
the probability density $|\Psi(r_n)|^2$
that the two nuclei are at contact, i.e.,
$\Lambda=A |\Psi(r_n)|^2$, where $A$ is the nuclear reaction constant
which is directly related to the astrophysical $S$ factor,
and $r_n$ is the nuclear radius. 
The relative wavefunction $\Psi$ of the two nuclei is governed
by the interatomic potential $V(r)$, which has been 
calculated in Paper I and II in the strong magnetic field regime. 
We can write $V(r)$ in the form:
$$V(r)={1\over r}-{1\over r_0}+E(\rH_2)-\Delta V(r),\eqno(B1)$$
where $r_0$ is the equilibrium separation,
$E(\rH_2)$ is the equilibrium energy of the molecule, and
the function $\Delta V(r)=0$ at $r=r_0$ and increases to 
$[E(\rH_2)-1/r_0-E({\rm He})]$ as $r$ decreases to zero ($E({\rm He})$
is the ground-state energy of He atom).  
For the molecule in the ground state, the relative energy $E_r$ 
of the two nuclei is simply $E(\rH_2)$ plus the zero-point energy 
$\hbar\omega_\parallel/2$ (we consider vibration along the
field axis only). In the WKB approximation, $|\Psi(r_n)|^2$ is
proportional to the WKB penetration factor $P$, i.e., 
$$
|\Psi(r_n)|^2 \propto P=\exp(-W),~~~~
W=2\sqrt{2\mu}
\int_{r_n}^{r_c}[V(r)-E_r]^{1/2} dr,
\eqno(B2)$$
where $r_c$ is the classical inner turning point,
$\mu$ is the reduced mass of the two nuclei in units of the
electron mass. 

Neglecting $\Delta V(r)$, equation (B2) can be easily integrated
to give $W\simeq \pi(2\mu r_c)^{1/2}$, with
$r_c=(1/r_0+\hbar\omega_\parallel/2)^{-1}$. A fitting 
formula for the numerical values of $r_0$ given in Paper I and Paper
II is $r_0\simeq 12.7(\ln b)^{-2.2}$. With $r_c\simeq r_0$ we have
$W\simeq 15.8\sqrt{\mu}(\ln b)^{-1.1}$.
The numerical results calculated using the exact potential $V(r)$ 
and zero-point energy (see Eq.~[2.4]) are given in Table 2,
where the values of $\lambda=W/\sqrt{\mu}$ for different field
strength are listed ($\lambda$ does not depend sensitively on $\mu$;
e.g., for $B_{12}=1$, $\lambda \simeq 1.74$ for pp molecule and
$\lambda \simeq 1.75$ for dd molecule.)
These numerical results can be fitted by
$$W=15.3\sqrt{\mu}\,(\ln b)^{-1.2},\eqno(B3)$$
which is reasonably close
to the approximate analytic expression. 
The factor $\lambda$ and the fusion rates for zero-field
molecules as calculated by Koonin \& Nauenberg (1989) are
also listed in Table 2 for comparison. 
The reactions considered are: p$+$p$\rightarrow ^2$H$+$e$^{+}+\nu_e$,
p$+$d$\rightarrow ^3$He$+\gamma$,
p$+$t$\rightarrow ^4$He$+\gamma$,
d$+$d$\rightarrow ^3$He$+$n$\oplus^3$H$+$p, 
d$+$t$\rightarrow ^4$He$+$n.
Rescale the zero-field fusion rates using the appropriate $\lambda$ 
for strong field, we immediately infer the
fusion rates for molecules in strong magnetic field.\footnote{$^7$}
{In the strong magnetic field, the wavefunction $\Psi$ for the two 
nuclei is not spherical, unlike the filed-free case.
Therefore the proportional factor in
$\Lambda \propto\exp(-\lambda\sqrt{\mu})$
is different for $B=0$ and for high field cases,
and our values for the fusion rate at high B are not exact. However,
since the dominant factor determining the fusion rate is
still the penetration factor, the error
should not exceed an order of magnitude.}
These are listed in Table 2.
Clearly, the strong magnetic field increases the fusion rate
by many orders of magnitude. The energy generation rate per gram 
of molecules is $\sim 10^{18}\Lambda$ erg$\,$g$^{-1}$s$^{-1}$.

\midinsert
\vfil
\leftskip 0.25in    
\rightskip 0.25in   
\newdimen\digitwidth
\setbox0=\hbox{\rm0}
\digitwidth=\wd0
\catcode`?=\active
\def?{\kern\digitwidth}
$$\vbox{
\tabskip=2em plus3em minus1.5em
\halign to4.5in{
\hfil#\hfil  &  \hfil#\hfil  &  \hfil#\hfil  &  \hfil#\hfil  &
\hfil#\hfil  &  \hfil#\hfil  &  \hfil#\hfil \cr
\multispan7\hfil {\bf TABLE 2} \hfil\cr
\noalign{\smallskip}
\multispan7\hfil{\bf Bound state fusion rates for isotopic hydrogen
molecules}\hfil\cr
\noalign{\medskip \hrule \vskip1pt \hrule \medskip}
  &Reactions$^{~a}$ &pp &pd &pt &dd &dt \cr
  &$\mu/m_p$ &$1/2$ &$2/3$ &$3/4$ &$1$ &$6/5$\cr
  &$\log_{10}A$ (cm$^3$s$^{-1}$)&$-39.1$&$-21.3$&
$-20.3$&$-15.8$&$-13.9$\cr
$B_{12}$&$\lambda^{~b}$&&&$\log _{10}\Lambda~{\rm s^{-1}}$&&\cr
\noalign{\medskip \hrule \medskip}
$0$&$4.13$&$-64.4$&$-55.0$&$-57.8$&$-63.5$&$-68.9$\cr
$1$&$1.74$&$-32.9$&$-18.7$&$-19.3$&$-19.0$&$-20.2$\cr
$5$&$1.35$&$-27.8$&$-12.7$&$-13.0$&$-11.8$&$-12.2$\cr
$10$&$1.22$&$-26.1$&$-10.8$&$-10.9$&$-9.3$&$-9.6$\cr
$100$&$.895$&$-21.8$&$-5.8$&$-5.7$&$-3.3$&$-3.0$\cr
$500$&$.740$&$-19.8$&$-3.5$&$-3.2$&$-0.4$&$0.2$\cr
\noalign{\medskip \hrule} 
}
}$$
\noindent{$^a$ The reactions are given in the text.}

\noindent{$^b$ $\lambda$ is defined via $P=\exp(-\lambda\sqrt{\mu})$;
and the values are given for the pd system; see text.}
\vskip 0.3in

\vfil
\endinsert

\bigskip
\centerline{\bf B.2 Internal Fusion of H$_N$ Chains}
\nobreak
\medskip

Larger H$_N$ molecules (before saturation; see \S 2.2) have similar 
electronic potential and excitation energies as H$_2$. Therefore
similar enhancement to the fusion rates also occurs in H$_N$
molecules, and this enhancement becomes more pronounced as $N$ (and 
the field strength) increases. For 
a long chain molecule H$_N$ with $1<<N<<N_s\sim 
[b/(\ln b)^2]^{1/5}$, the spacing $r_o$ along a field line 
between adjacent protons decreases with increasing $N$ approximately 
as\footnote{$^8$}{Note that this is the asymptotic limiting result.
For small $N$, the scaling with $b$ is closer to 
$r_0\sim (\ln b)^{-2}$.} 
$r_0\sim 1/(N^2\ln b)$ (Paper I), and thus $W\sim r_0^{1/2}\sim 1/N$.
A deviation $\delta r$
from the equilibrium spacing would give an excess potential
$\delta V\sim (\ln b)(\delta r)^2/r_0^3$. Thus  
the fractional zero-point vibration amplitude $\Delta r/r_0$ 
is of order $(m_e/m_p)^{1/4}N^{1/2}$, i.e., the aligned vibrations
become more pronounced as $N$ increases. Both the decrease in $r_0$
and the increase in $\Delta r/r_0$ tend to enhance the nuclear
reaction rate. 

\bigskip
\centerline{\bf B.3 Fusion in Metallic Hydrogen}
\nobreak
\medskip

As mentioned in \S 2.3, the zero-pressure metallic hydrogen 
has nonuniform electron distribution. Although this screening
effect will certainly increase the fusion rate, the essential 
feature can be obtained in the uniform electron gas model. 
As an estimate, we can simply use the fitting formula of Salpeter and
van Horn (1969; see also Schramm \& Koonin 1990) for the pycnonuclear
reaction rate, except that the zero-pressure density is given by
equation (3.1). For pd reaction, the rate per deuteron is
$\Lambda_{\rm pd}\simeq 2\times
10^{13}\rho_6^{7/12}\!\exp(-21.8\rho_6^{-1/6})$
~s$^{-1}$. Thus a deuteron embedded in a cold hydrogen plasma
has lifetime $\Lambda_{\rm pd}^{-1}= 10^{9.6}$ years at $B_{12}=2$
and one year at $B_{12}=13$. 
For pp fusion, the reaction rate is $\Lambda_{\rm pp}\simeq 800\,
\rho_6^{7/12}\!\exp(-18.89\rho_6^{-1/6})$ yr$^{-1}$.
At $B_{12}=60,\,100,\,500$, corresponding to zero-pressure
densities $\rho_{s,0}\simeq 7.6\times 10^4,\,1.4\times 10^5,\,10^6$
g~cm$^{-3}$, we find 
$\Lambda_{\rm pp}^{-1}\simeq 10^{10.3},\,10^9,\,10^{5.3}$ yrs,
respectively. Thus for non-accreting neutron star, 
there is an upper limit to the field strength 
above which the surface hydrogen layer is absent.
Comparing with Table 2, we see that 
at a given field strength, the fusion rate is larger 
in the metallic state than in molecule. The reason is
that the mean ion separation in the condensed state decreases 
with increasing field strength as a power-law function of
$B$, whereas the mean interatomic separation 
in a molecule decreases only logarithmically. 
Also note that as the matter density or temperature increase, 
the magnetic field effects
become less important because many Landau levels are
occupied by the electrons. 



\vfill\eject
\bigskip
\bigskip
\centerline{\bf REFERENCES}
\nobreak
\bigskip
\def\bysame{\hbox to 50pt{\leaders\hrule height 2.4pt depth -2pt\hfill .\ }}

\hi{
Abrahams, A.~M., \& Shapiro, S.~L. 1991, ApJ, 374, 652}





\hi{
Ashcroft, N.~W., \& Mermin, N.~D. 1976, Solid State Physics (Saunders
College: Philadelphia), p.337}

\hi{
Avron, J.~E., Herbst, I.~B., \& Simon, B. 1978, Ann. Phys. (NY), 114,
431}

\hi{
Becker, W. 1995, Ann. N.Y. Acad. Sci., 759, 250}


\hi{
Blaes, O, \& Madau, P. 1993, ApJ, 403, 690}


\hi{
Brinkmann, W. 1980, A\&A, 82, 352}

\hi{
Callaway, J., \& March, N.~H. 1984, Solid State Phys., 38, 135}


\hi{
Canuto, V., Lodenquai, J., \& Ruderman, M. 1971, Phys. Rev., D3, 2303}

\hi{
Ceperley. D., \& Alder, B. 1980, Phys. Rev. Lett., 45, 566}



\hi{
Chiu, H.~Y., \& Salpeter, E.~E. 1964, Phys. Rev. Lett., 12, 413}

\hi{
Duncan, R.~C., \& Thompson, C. 1992, ApJ, 392, L9}

\hi{
Edelstein, J., \& Bowyer, S. 1996, in Pulsars: Problems and Progress,
Ed. S. Johnston, M.~A. Walker \& M. Bailes (ASP Conference Series,
Vol.~105), p291.


\hi{
Flowers, E.~G., et al.
1977, ApJ, 215, 291}

\hi{
Fushiki, I., Gudmundsson, E.~H., \& Pethick, C.~J. 1989, ApJ, 342, 958}

\hi{
Fushiki, I., Gudmundsson, E.~H., Pethick, C.~J., \& Yngvason, J. 1992, 
Ann. Phys. (NY), 216, 29}

\hi{
Gnedin, Yu.~N., Pavlov, G.~G., \& Tsygan, A.~I. 1974,
Sov. Phys. JETP, 39, 201}





\hi{
Hansen, J.~P., \& Pollock, E.~L. 1973, Phys. Rev. A8, 3110}

\hi{
Hernquist, L. 1985, MNRAS, 213, 313}

\hi{
Herold, H., Ruder, H., \& Wunner, J. 1981, J.~Phys. B: At. Mol. Phys., 
14, 751}

\hi{
Horing, N.~J. 1969, Ann. Phys. (NY), 54, 405}


\hi{
Itoh, N. 1975, MNRAS, 173, 1P}


 
\hi{
Jones, P.B. 1985, MNRAS, 216, 503}

\hi{
\bysame 1986, MNRAS, 218, 477}

\hi{
Kadomtsev, B.B. 1970 Sov. Phys. JETP, 31, 945}
 
\hi{
Khersonskii, V.~K. 1987, Sov. Astron., 31, 225}

\hi{
Koonin, S.~E., \& Nauenberg, M. 1989, Nature, 339, 691}

\hi{
Lai, D., Salpeter, E.~E., \& Shapiro, S.~L. 1992, Phys. Rev. A, 45,
4832 (Paper I)}


\hi{
Lai, D., \& Salpeter, E.~E. 1996, Phys. Rev. A, 53, 152.
(Paper II)}

\hi{
\bysame 1995, Phys. Rev. A, 52, 2611 (Paper III)}




\hi{
Landau, L.~D., \& Lifshitz, E.~M. 1980, Statistical Physics, Part 1, 
3rd Ed. (Pergamon: Oxford)}

\hi{
Lieb, E.~H., Solovej, J.~P., \& Yngvason, J. 1994,
Commun. Math. Phys., 161, 77}


\hi{
Lodenquai, J., Canuto, V., Ruderman, M., \& Tsuruta, S. 1974, ApJ,
190, 141}


\hi{
Miller, M.C. 1992, MNRAS, 255, 129}

\hi{
Nelson, R.~W., Salpeter, E.~E., \& Wasserman, I. 1993, ApJ, 418, 874}

\hi{
Neuhauser, D., Koonin, S.~E., \& Langanke, K. 1987, Phys. Rev. A, 36,
4163}



\hi{
Paczy\'nski, B. 1992, Acta Astron., 42, 145}


\hi{
Pavlov, G.~G., Shibanov, Y.~A., Ventura, J., \& Zavlin, V.~E.
1994, A\&A, 289, 837}

\hi{
Pavlov, G.~G., Shibanov, Y.~A., Zavlin, V.~E., \& Meyer, R.~D. 1995, 
in ``Lives of Neutron Stars'', eds. A. Alpar, U. Kiziloglu \& 
J. van Paradijs (Kluwer: Dordrecht)}

\hi{
Pavlov, G.~G., Zavlin, V.~E., Tr\"umper, J., \& Neuh\"auser, R.
1996, ApJ, 472, L33}

\hi{
Pethick, C.~J. 1992, Rev. Mod. Phys., 64, 1133}

\hi{
Potekhin, A.~Y. 1994, J. Phys. B., 27, 1073}

\hi{
Potekhin, A.~Y., \& Pavlov, G.~G. 1993, ApJ, 407, 330}

\hi{
Rajagopal, M., \& Romani, R.~W. 1996, ApJ, 461, 327}

\hi{
Reisenegger, A. 1995, ApJ, 442, 749}

\hi{
Romani, R.~W. 1987, ApJ, 313, 718}


\hi{
Ruder, H. et al. 1994, Atoms in Strong Magnetic Fields
(Springer-Verlag)}

\hi{
Ruderman, M. 1971, Phys. Rev. Lett., 27, 1306}

\hi{
\bysame 1974, in ``Physics of Dense Matter'' (I.A.U.~Symp.~No.~53),
ed.~C.J. Hansen (Dordrecht-Holland: Boston), p.117}


\hi{
Salpeter, E.~E., \& van Horn, H.~M. 1969, ApJ, 155, 183}

\hi{
Schramm, S., \& Koonin, S.~E. 1990, ApJ, 365, 296}



\hi{
Shemi, A. 1995, MNRAS, 275, L7}

\hi{
Silant'ev, N.~A., \& Yakovlev, D.~G. 1980, Ap. Space Sci., 71, 45}

\hi{
Thompson, C., \& Duncan, R.~C. 1993, ApJ, 408, 194}

\hi{
Treves, A., \& Colpi, M. 1991, A\&A, 241, 107}

\hi{
Tsuruta, S. 1964, Ph.D. thesis, Columbia University}

\hi{
Umeda, H., Tsuruta, S., \& Nomoto, K. 1994, ApJ, 433, 256}


\hi{
Ventura, J., Herold, H., Ruder, H., \& Geyer, F. 1992, A\&A, 261, 235}



\hi{
Walter, F.~M., Wolk, S.~J., \& Neuh\"auser, R. 1996, Nature, 379, 233}


\hi{
Zel'dolvich, Ya.~B., \& Gershtein, S.~S. 1961, Sov. Phys. Uspekhi, 3, 593}

\hi{
Zavlin, V.~E. et al. 1995, A\&A, 297, 441}

\hi{
Zavlin, V.~E., Pavlov, G.~G., \& Shibanov, Y.~A. 1996, A\&A, 315, 141}

\vfill\eject
\bigskip
\centerline{\bf FIGURE CAPTIONS}
\nobreak
\bigskip

\medskip\noindent
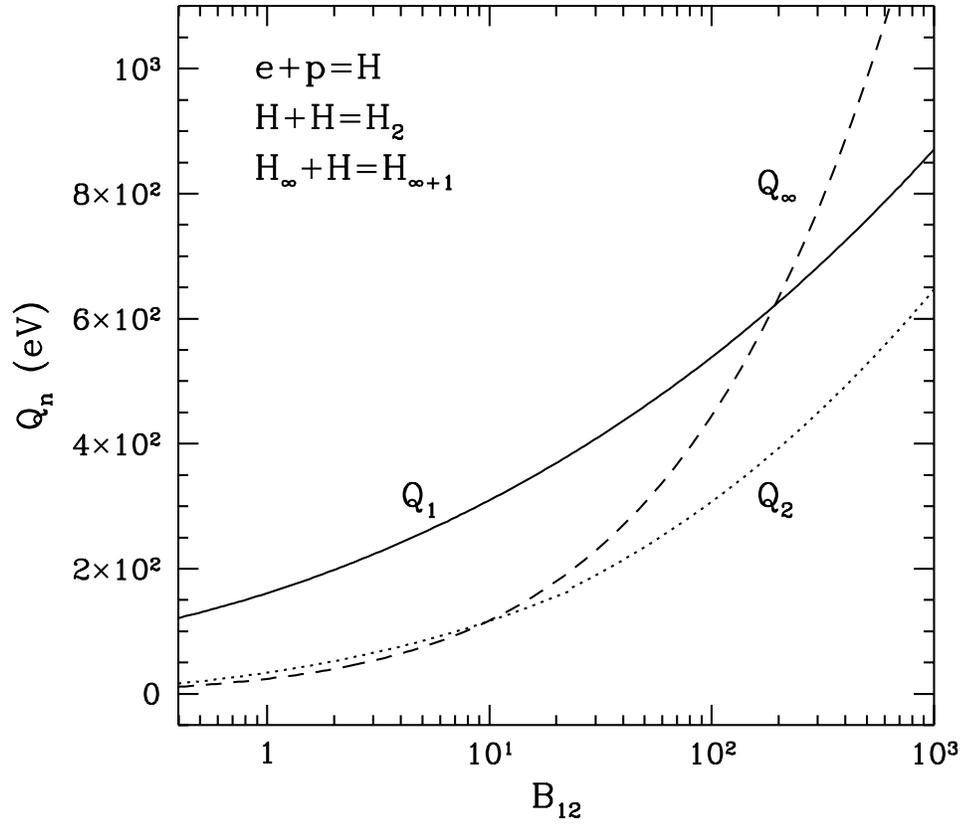
{\bf FIG.~1}.---
Energy releases from several atomic and molecular processes as a function 
of the magnetic field strength.
The solid line shows the ionization energy $Q_1$ of H atom, 
the dotted line shows the dissociation energy $Q_2$ of H$_2$, and the 
dashed line shows the cohesive energy $Q_\infty$ of linear chain
H$_\infty$. The zero-point energy corrections have been included in
$Q_2$ and $Q_\infty$. 

\medskip\noindent
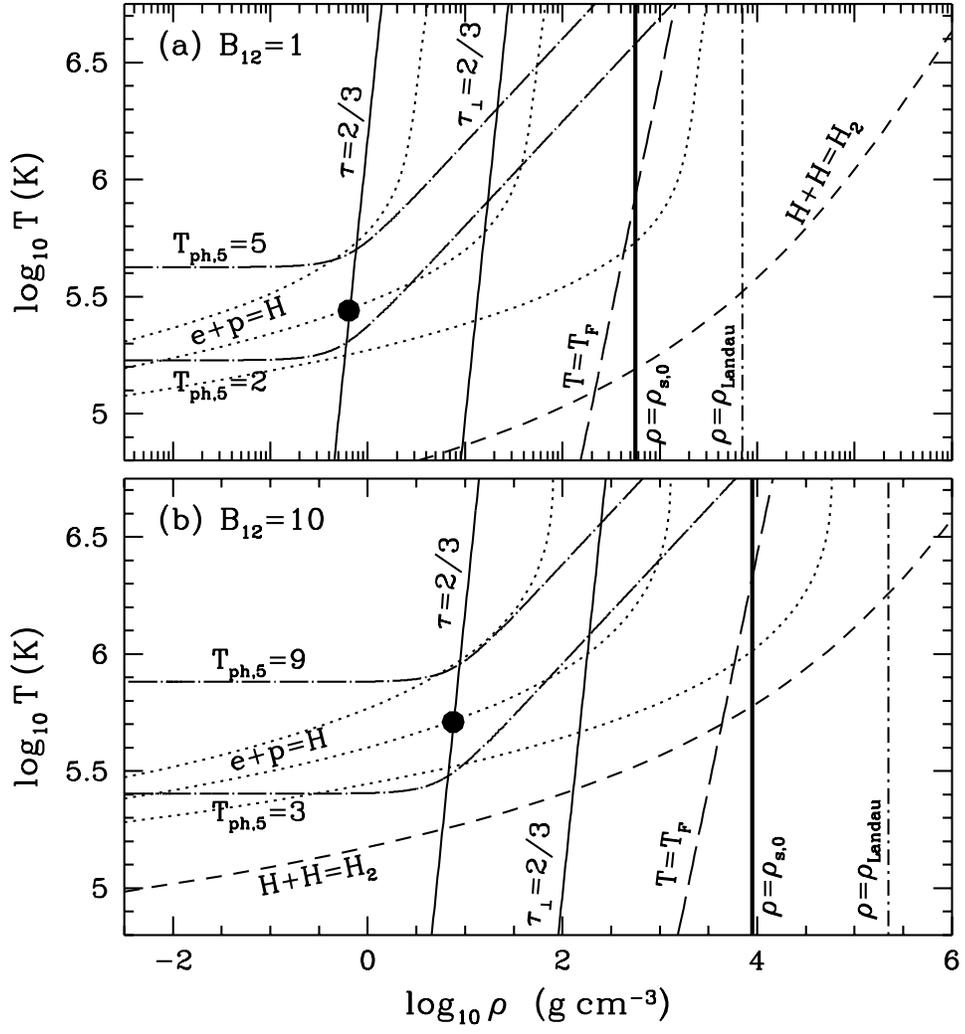
{\bf FIG.~2}.---
Temperature-density diagram of the atmosphere of a neutron star
with surface field strength (a) $B_{12}=1$ and (b) $B_{12}=10$, 
both with $g_{14}=1$. 
The solid lines correspond to the photosphere (``$\tau\simeq 2/3$'') 
and the extraordinary photon emission 
region (``$\tau_\perp\simeq 2/3$''); 
the dotted curves (``e+p=H'') correspond to constant neutral H
fractions, with $n(\rH)/n_g=0.9$ for the lower curves, 
$n(\rH)/n_g=0.5$ (half/half ionization) for the 
middle curves, and $n(\rH)/n_g=0.1$ for the upper curves; 
the dashed curves (``H+H=H$_2$'') correspond to the half/half
dissociation of H$_2$; the long-dashed lines (``$T=T_F$'') show the
Fermi temperature; the dot-long-dashed lines show selected
temperature-density profiles of atmospheres with different
photospheric temperatures ($T_{\rm ph,5}=5,~2$ for $B_{12}=1$ and
$T_{\rm ph,5}=9,~3$ for $B_{12}=10$); the dot-dashed vertical lines 
(``$\rho=\rho_{\rm Landau}$'') correspond to the density above which
the electrons occupy the excited Landau level; the dark vertical lines
correspond to the condensation density $\rho_{s,0}$; 
The filled circles correspond to the photospheric ``ionization
temperature'' $T_{\rm ph,ir}$ (see text).

\medskip\noindent
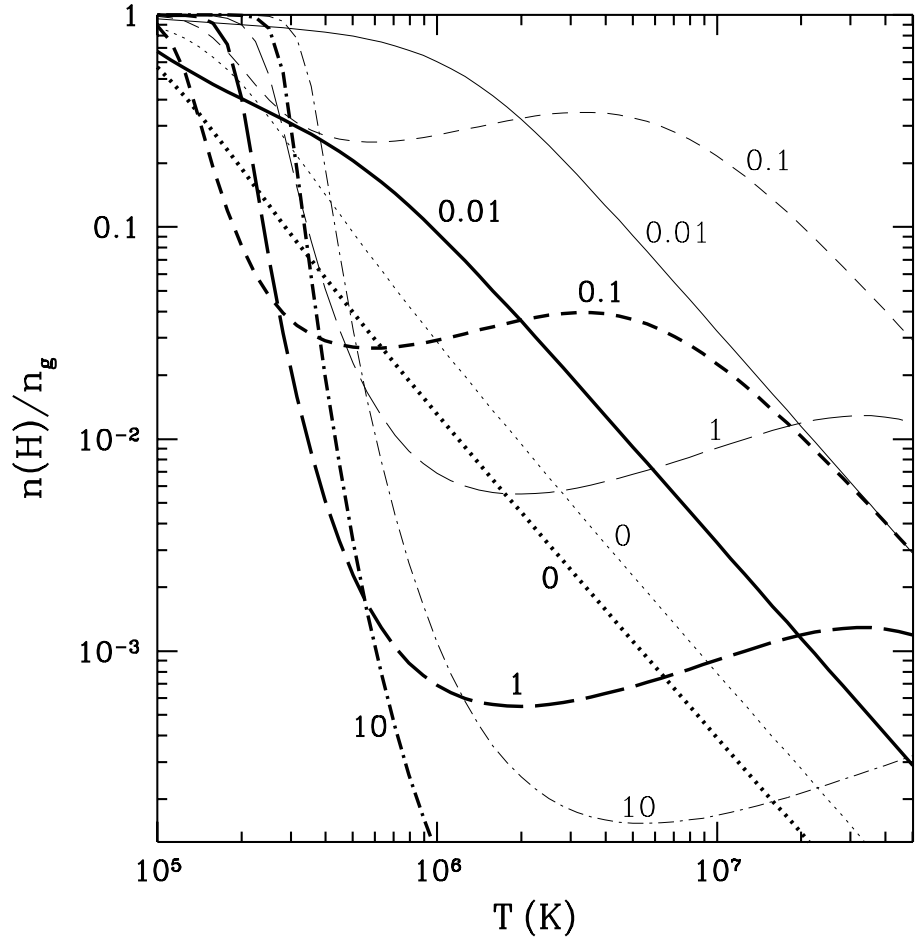
{\bf FIG.~3}.---
Neutral fraction $n(\rH)/n_g$ as a function of temperature 
at a fixed density $\rho=n_g m_p=0.01$ g/cm$^3$ (heavy lines) 
and $0.1$ g/cm$^3$ (light lines).
The curves are labeled by the field strength:
$B_{12}=0.01$ (solid lines), $0.1$
(short-dashed lines), $1$ (long-dashed lines), $10$ (dot-dashed
lines). The zero-field results are also shown for comparison
(dotted lines). 


\medskip\noindent
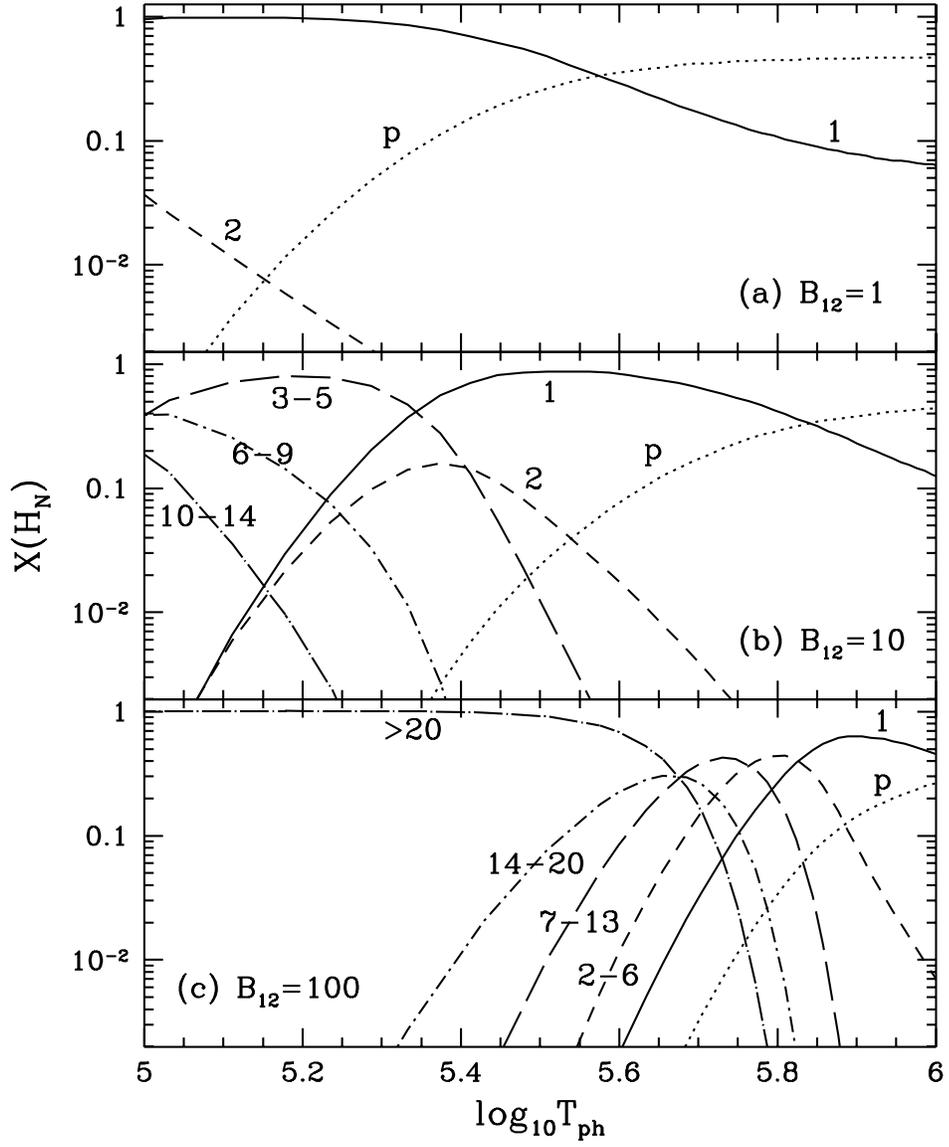
{\bf FIG.~4}.---
Abundance of various H species as a function 
of the photosphere temperature at a point with $P=P_{\rm ph}/2$
(just a little above the photosphere) for three field strengths
(a)$~B_{12}=1$, (b)$~10$, and (c)$~100$. The fractional
number density are defined by
$X(\rH_N)=n(\rH_N)/n_{\rm tot}$, $X_p=n_p/n_{\rm tot}$, with 
$n_{\rm tot}=2n_p+n(\rH)+n(\rH_2)+\cdots$. 
The numbers labeling each curve
specify $N$, with, e.g., ``3-5'' indicating $X(\rH_3)+X(\rH_4)
+X(\rH_5)$. 

\medskip\noindent
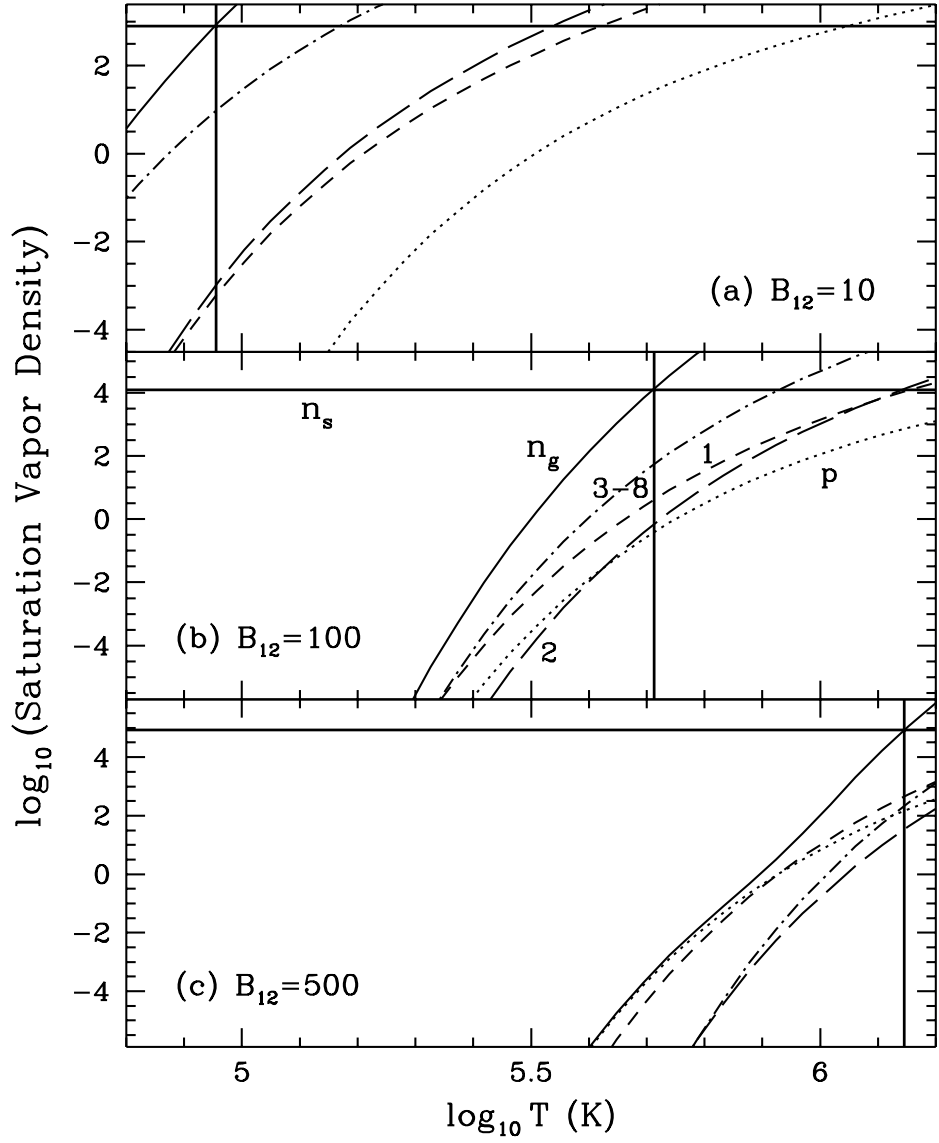
{\bf FIG.~5}.---
The saturation vapor densities of various species 
(in the atomic units, $a_o^{-3}$) of condensed metallic hydrogen 
as a function of temperature for different magnetic 
field strengths: (a) $B_{12}=10$; (b) $B_{12}=100$; (c) $B_{12}=500$.
The dotted curves give $n_p$, the short-dashed curves give $n(\rH)$,
the long-dashed curves give $n(\rH_2)$, the dot-dashed curves
give $[3n(\rH_3)+4n(\rH_4)+\cdots+8n(\rH_{8})]$, and the
solid curves give the total baryon number density in the vapor
$n_g=n_p+n(\rH)+2n(\rH_2)+\cdots$. 
The horizontal solid lines denote the condensation density $n_s\simeq
50\,B_{12}^{6/5}$ (a.u.), while the vertical solid lines correspond to 
the critical condensation temperature at which $n_g=n_s$.

\medskip\noindent
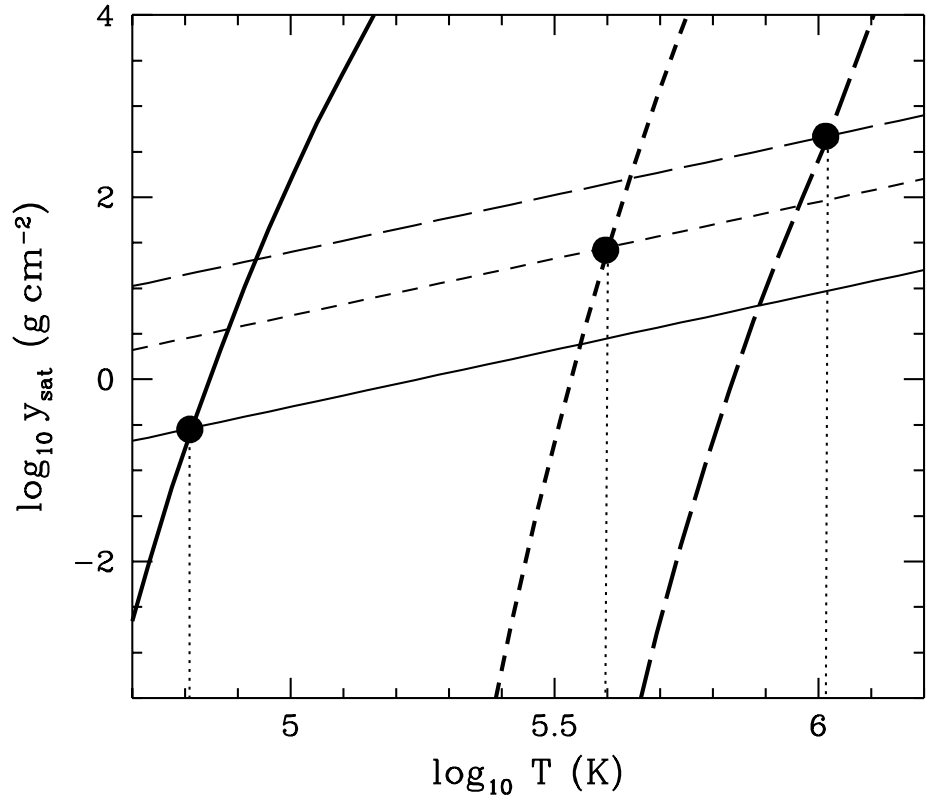
{\bf FIG.~6}.---
Column density of saturated vapor. 
The heavy (steeper) lines show the column density $y$ of the
nondegenerate gas above the condensed metallic phase as 
a function of the temperature at the phase boundary, 
the lighter lines show the threshold 
column densities $y_{\rm th}$ (Eq.~[4.10]) below which the vapor 
is optically thin to free-free absorption. 
The solid lines are for $B_{12}=10$, 
the short-dashed lines for $B_{12}=100$, and the long-dashed lines 
for $B_{12}=500$, all with $g_{14}=1$. 
The filled circles and vertical dotted lines
correspond to the threshold photospheric temperature
$T_{\rm ph,th}$ below which the saturated vapor is optically thin.

\end